\documentclass[11pt]{article}
\usepackage[a4paper,margin=2.3cm]{geometry}
\usepackage[T1]{fontenc}
\usepackage{lmodern}
\ifdefined\pdfgentounicode\input{glyphtounicode}\pdfgentounicode=1\fi
\usepackage{amsmath,amssymb}
\usepackage{graphicx}
\usepackage{booktabs}
\usepackage[super,sort&compress,comma]{natbib}
\usepackage[hidelinks]{hyperref}

\makeatletter\renewcommand\@biblabel[1]{#1.}\makeatother
\AtBeginDocument{\setlength{\parindent}{0pt}\setlength{\parskip}{0.6ex}}
\begin{document}
\title{\bfseries Transformer-inspired architectures make machine-learned electron densities accurate and transferable enough to bypass self-consistency}
\author{Jeheon Woo\\[2pt] \small Center for AI and Natural Sciences, Korea Institute for Advanced Study, Seoul 02455, Republic of Korea\\ \small\texttt{jhwoo1905@kias.re.kr}}
\date{}
\maketitle
\begin{abstract}
We introduce machine learning models that accurately predict the electron density, the central variable of density functional theory. They are built on Transformer-inspired architectures, which require little computation per parameter and can therefore be enlarged at low cost. The proposed models, SwinUNETR-V2 and its attention-free counterpart NeXtUNETR, achieve state-of-the-art accuracy on QM9, with normalized mean absolute errors in the density averaging 0.075\% and 0.046\%, respectively. The predicted densities are accurate enough to bypass the self-consistent iteration: a single Kohn--Sham step reproduces the self-consistent electronic structure. For NeXtUNETR, the median errors are 0.036~meV in the total energy, 0.65~meV~\AA{}$^{-1}$ in the forces and 1.8~meV in the HOMO--LUMO gap. The forces not only recover the self-consistent minima in geometry optimization but also successfully drive ab initio molecular dynamics. Fine-tuned on small datasets, both networks also transfer accurately to systems containing elements absent from QM9 or larger than its molecules.
\end{abstract}

Kohn--Sham density functional theory\cite{HK1964,KS1965} is the most widely used electronic-structure method in chemistry and materials science, and the electron density is its central variable: it determines all ground-state properties\cite{HK1964}, and the Kohn--Sham Hamiltonian is built from it. Finding the density is the costly part: the self-consistent-field (SCF) iteration diagonalizes a Hamiltonian built from a trial density and updates the density until the input and output densities agree. Because this iteration is repeated at every step of quantum-chemical simulations such as geometry optimization and molecular dynamics, it dominates their cost. Machine learning offers two ways around it. Machine-learned interatomic potentials (MLIPs) avoid the electronic problem by mapping atomic positions directly to energies and forces\cite{BehlerParrinello07,MACE,Unke2021}, and foundation models now span most of the periodic table\cite{MACE-MP-0}, but they return no electronic structure. Learning the density\cite{Brockherde2017,Li2025}, or the density matrix or Hamiltonian\cite{DeepH,Shao2023}, retains it: from a predicted density, a single Kohn--Sham step returns energies, forces and orbitals together, and their quality is set by that of the density.

The density at a point is determined mainly by the atoms nearby, a property known as the nearsightedness of electronic matter\cite{Prodan2005}, so a model can learn it from local environments and apply it to larger systems. A useful model must be accurate, because density errors propagate into every derived property, and transferable, because a model trained on small molecules must also serve larger, charged or differently composed systems. Most density models describe a molecule by its atoms: kernel methods and graph neural networks, mostly rotationally equivariant, expand the density in atom-centered or floating basis functions or predict it point by point\cite{Grisafi2019,DeepDFT,ChargE3Net,SCDP,ELECTRA}. Others work on the real-space grid of the calculation, where the density forms a three-dimensional image, and refine a superposition of atomic densities (SAD), alone\cite{Li2025} or together with grid-projected atomic fingerprints\cite{DeepSCF}, with convolutional neural networks (CNNs). Li et al. cast this refinement as image super-resolution\cite{Li2025}: their residual network (here SR-ResNet), trained on QM9 (ref.~\citenum{QM9}), was more accurate than earlier models and transferred to unseen geometries and elements after fine-tuning on small datasets.

If density prediction is an imaging problem, advances in computer vision may carry over to it. The Transformer\cite{Transformer}, developed for natural language processing, was applied to images by the Vision Transformer\cite{ViT}, and hierarchical variants such as the Swin Transformer\cite{Swin}, with self-attention in local windows and a hierarchy of resolutions, reach or exceed the accuracy of CNNs such as ResNet\cite{ResNet} on major vision benchmarks. In three dimensions, SwinUNETR and SwinUNETR-V2 (refs.~\citenum{SwinUNETR,SwinUNETR-V2}) are widely used for volumetric medical image segmentation, where the convolutional 3D U-Net\cite{3DU-Net} had long been the standard. Transformers have also reshaped CNNs: ConvNeXt\cite{ConvNeXt} reached the accuracy of Swin by adopting its macro-design, with large-kernel depthwise convolutions in place of attention, and 3D UX-Net and MedNeXt carried this design to volumetric images\cite{3DUX-Net,MedNeXt}. We call both kinds of networks Transformer-inspired: they share the hierarchical macro-design and block recipe of the Swin Transformer, with or without attention. To our knowledge, density refinement on real-space grids treated as images has so far used only CNNs\cite{DeepSCF,Li2025}, and whether Transformer-inspired architectures make densities more accurate and transferable has not been tested.

An accurate density can replace the SCF iteration altogether. The Kohn--Sham Hamiltonian built from the predicted density is diagonalized once, and this single Kohn--Sham step gives the orbitals, a new density and the energy. Because the Kohn--Sham energy is stationary at the self-consistent density, the error of this one-step energy is second order in the density error\cite{Harris1985,FH1989}. Forces need more care. The predicted density is not exactly self-consistent, so, as for other non-self-consistent energies\cite{Niklasson2017,Niklasson2023}, the exact gradient of the one-step energy retains a term, omitted by standard Kohn--Sham forces, from the response of the density to the nuclear positions. For a learned density, the response must be propagated through the network; we call the added term the non-self-consistency correction (NSCC). Non-self-consistent properties of learned densities have so far been tested on fixed structures\cite{Li2025,DeepSCF,ChargE3Net,Focassio2023,Grisafi2023}, and SCF-free dynamics has used system-specific models of the density\cite{Brockherde2017,Hazra2025} or density matrix\cite{Shao2023,Hazra2024}. Whether a transferable density model makes one Kohn--Sham step accurate enough for geometry optimization and energy-conserving molecular dynamics has remained open.

Here we show that Transformer-inspired architectures make learned densities accurate and transferable enough to bypass the SCF iteration. We adapt two U-shaped networks to refine the SAD guess: SwinUNETR-V2, with windowed self-attention, and NeXtUNETR, an attention-free counterpart with the same hierarchical design. Trained on QM9, they reach normalized mean absolute errors (NMAEs) in the density averaging 0.075\% and 0.046\%, respectively, below those of any published model trained on the same data. At equal training data, NeXtUNETR is the most accurate network at every size tested, and in the same U-shaped network, ConvNeXt blocks are markedly more accurate than Swin blocks. One Kohn--Sham step from the NeXtUNETR densities gives median errors of 0.036~meV in the total energy and 0.65~meV~\AA{}$^{-1}$ in the forces, and both networks give frontier orbital energies within a few meV. With or without the NSCC, the forces relax 200 molecules from nearby starting structures, 97--98.5\% of them to within 0.1~\AA{} of the SCF minima. The NSCC, which also lowers the force error for most molecules, makes molecular dynamics energy-conserving for a molecule with more heavy atoms than any training molecule. Both networks also transfer: without fine-tuning, SwinUNETR-V2 is the most accurate model on charged fragments with unseen elements and on water clusters, and fine-tuned on about a hundred structures, both outperform SR-ResNet on these fragments. Architectural advances made for images thus carry over to electronic structure.

\section*{Results}
\subsection*{Density refinement and the one-step calculation}

A SAD guess density is sampled on the real-space grid of the calculation, and a neural network returns the refined density $\rho_\mathrm{ML}$ on the same grid (Fig. 1a). The Kohn--Sham Hamiltonian matrix $H[\rho_\mathrm{ML}]$ is built and diagonalized once. Like a converged Kohn--Sham calculation, this single step gives the orbitals and orbital energies, the density matrix and the properties derived from it, and the one-step total energy, whose analytic derivative gives the forces. We report forces with the NSCC (w/ NSCC), which are the exact gradient of the one-step energy, and without it (w/o NSCC), the standard Kohn--Sham force.

\begin{figure}[!tbp]
{\centering
\makebox[\linewidth][c]{\includegraphics[width=180.0mm,height=0.55\textheight,keepaspectratio]{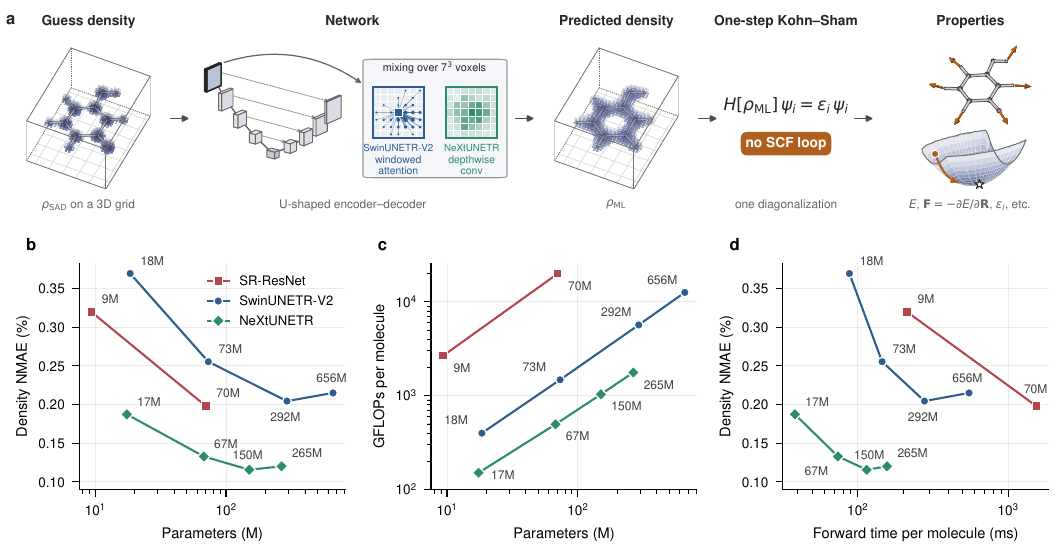}}\par}
\medskip\small
\textbf{Fig. 1 | One-step Kohn--Sham calculation from predicted densities, and the architectures compared.} \textbf{a}, Schematic workflow. A U-shaped network refines the superposition of atomic densities (SAD) on the real-space grid (left) into the predicted density $\rho_\mathrm{ML}$. Its encoder blocks mix features within 7 $\times$ 7 $\times$ 7-voxel regions (enlarged): SwinUNETR-V2 by windowed attention, NeXtUNETR by a depthwise convolution. One diagonalization of the Kohn--Sham Hamiltonian $H[\rho_\mathrm{ML}]$, without a self-consistent-field (SCF) loop, gives the energy, forces and orbital energies (right: the one-step energy surface). \textbf{b}, Density NMAE against parameter count for three families trained on the same 8,000 QM9 molecules (QM9-subset-8k). \textbf{c}, Floating-point operations per molecule against parameter count. \textbf{d}, Density NMAE against network forward time per molecule (one NVIDIA H200 NVL, batch size 1, fp32). \textbf{b}--\textbf{d}, Means over 500 randomly sampled QM9 test molecules; one training run per model; labels give the parameter count; counting and timing in Methods. Extended Data Fig. 1 shows the details of the architectures.

\end{figure}

Both networks are U-shaped encoder--decoders that process the grid at its input resolution and five successively halved ones (Extended Data Fig. 1). SwinUNETR-V2 (ref.~\citenum{SwinUNETR-V2}) mixes information within 7 $\times$ 7 $\times$ 7-voxel windows by self-attention, whose weights depend on the input; NeXtUNETR mixes a neighborhood of the same size by a depthwise convolution\cite{ConvNeXt}, whose weights are fixed after training. The baseline is the SR-ResNet of Li et al.\cite{Li2025}.

All densities are Perdew--Burke--Ernzerhof (PBE) valence densities (Methods). We measure the error of a density by its NMAE over the grid, in percent of the electron count (Methods). The energy error is $\varepsilon_E = |E_\text{1-step} - E_\mathrm{SCF}|$ per molecule, and the force error $\varepsilon_F$ is the absolute error of a Cartesian force component, summarized by its mean or median over all components.

\subsection*{Architecture comparison at equal training data}

We trained ten models of the three families on the same 8,000 QM9 molecules\cite{QM9} (QM9-subset-8k) and evaluated them on 500 molecules randomly sampled from the QM9 test split (Fig. 1b--d). The U-shaped networks were enlarged through the channel width $C$, the number of feature channels in the first encoder stage, which sets the widths of all later stages; SR-ResNet was enlarged in width and depth. NeXtUNETR is the most accurate family at every size. Its smallest model, with 17 million parameters and an NMAE of 0.187\%, is more accurate than every SwinUNETR-V2 model (best 0.204\%, 292 million parameters) and the largest SR-ResNet (0.198\%, 70 million). At the same channel width, $C = 96$, NeXtUNETR-67M reaches 0.133\% against 0.204\% for SwinUNETR-V2-292M, with a quarter of the parameters and one-eleventh of the floating-point operations.

In every family, the floating-point operations grow in proportion to the parameter count (Fig. 1c), but the operations per parameter differ: 7--9 GFLOPs per million parameters for NeXtUNETR, 19--22 for SwinUNETR-V2 and 285 for SR-ResNet, because the U-shaped networks place most of their parameters on coarse grids, whereas SR-ResNet runs all its residual blocks at half resolution. The U-shaped networks can therefore be enlarged at little computational cost, and the added capacity makes them more accurate: in both families the NMAE falls with model size until it levels off at 150--300 million parameters (Fig. 1b).

In ablations (Extended Data Table 1), the encoder block makes a large difference. With Swin Transformer blocks in place of the ConvNeXt blocks of its encoder, NeXtUNETR's NMAE rises from 0.133 to 0.250\%. The two blocks have the same feed-forward layers and differ chiefly in how they mix space, by windowed attention or by depthwise convolution, and in their normalization (layer or group). Removing the convolutional residual blocks on the downsampled skips and in the bottleneck, which SwinUNETR-V2 inherits from SwinUNETR\cite{SwinUNETR}, nearly halves its parameter count, from 292 to 154 million, but raises the NMAE only from 0.204 to 0.209\%.

For the remaining results, SwinUNETR-V2 and NeXtUNETR were trained on the full QM9 training set (123,835 molecules) in the SwinUNETR-V2-292M and NeXtUNETR-67M configurations, which share $C = 96$. SR-ResNet is Li et al.'s published 9-million-parameter QM9 model.

\subsection*{Densities, energies and orbitals on QM9}

On the 10,000 QM9 test molecules, the three models reduce the mean NMAE of the SAD guess, 13.9\%, to 0.138\% for SR-ResNet, 0.075\% for SwinUNETR-V2 and 0.046\% for NeXtUNETR (Table 1). The last two are 46\% and 67\% below that of SR-ResNet and below the 0.116\% reported for BOA, an equivariant basis-function network trained on the same densities and split\cite{BOA}. The ranking of the three models holds for almost every molecule.

\begin{table}[!tbp]
\small
\textbf{Table 1 | Accuracy on the QM9 test set.} Density NMAE of the predicted density and of the density after one Kohn--Sham step, and errors of the one-step energies and forces (${\Delta}$HOMO, ${\Delta}$LUMO and ${\Delta}$gap: absolute errors of the highest occupied and lowest unoccupied molecular orbital (HOMO and LUMO) energies and of the gap). For the SAD guess and the three models evaluated here, mean and median over the 10,000 test molecules ($\varepsilon_F$: over 539,613 force components); in parentheses, the mean without the two molecules with a reference HOMO--LUMO gap below 1~eV\textsuperscript{a}. For the other models, mean errors as reported. Best value in bold.

\begin{center}
\footnotesize
\setlength{\tabcolsep}{4pt}
\resizebox{\ifdim\width>\linewidth\linewidth\else\width\fi}{!}{%
\begin{tabular}{llrrrrrrrr}
\toprule
\multicolumn{2}{c}{} & \multicolumn{2}{c}{Density NMAE (\%)} & \multicolumn{4}{c}{One-step energies (meV)} & \multicolumn{2}{c}{One-step forces $\varepsilon_F$ (meV~\AA{}$^{-1}$)} \\
\cmidrule(lr){3-4}\cmidrule(lr){5-8}\cmidrule(lr){9-10}
Model &  & Predicted & One step & $\varepsilon_E$ & ${\Delta}$HOMO & ${\Delta}$LUMO & ${\Delta}$gap & w/o NSCC & w/ NSCC \\
\midrule
SAD guess (input) & mean & 13.9 & -- & -- & -- & -- & -- & -- & -- \\
 & median & 14.0 & -- & -- & -- & -- & -- & -- & -- \\
\multicolumn{10}{l}{\textbf{Previous work}} \\
eqDeepDFT (ref.~\citenum{DeepDFT}) & mean & 0.27\textsuperscript{b} & -- & -- & -- & -- & -- & -- & -- \\
ELECTRA (ref.~\citenum{ELECTRA}) & mean & 0.176\textsuperscript{b} & -- & -- & -- & -- & -- & -- & -- \\
BOA (ref.~\citenum{BOA}) & mean & 0.116; 0.134\textsuperscript{b} & -- & -- & -- & -- & -- & -- & -- \\
SR-ResNet (ref.~\citenum{Li2025})\textsuperscript{c} & mean & 0.138 & 0.058 & 2.84 (1.03) & 11.7 & 9.4 & 10.7 & 11.87 (11.63) & 17.05 (8.68) \\
 & median & 0.129 & 0.050 & 0.52 & 9.2 & 6.4 & 7.1 & 5.54 & 3.37 \\
\multicolumn{10}{l}{\textbf{This work}} \\
SwinUNETR-V2 & mean & 0.075 & 0.030 & 0.90 (0.26) & 6.1 & 4.5 & 5.7 & 6.21 (6.04) & 10.12 (4.36) \\
 & median & 0.070 & 0.026 & 0.14 & 4.8 & 3.0 & 3.7 & 2.93 & 2.37 \\
NeXtUNETR & mean & \textbf{0.046} & \textbf{0.015} & \textbf{0.42 (0.087)} & \textbf{2.8} & \textbf{2.2} & \textbf{2.9} & \textbf{3.21 (3.09)} & \textbf{4.46 (1.63)} \\
 & median & \textbf{0.043} & \textbf{0.013} & \textbf{0.036} & \textbf{2.1} & \textbf{1.4} & \textbf{1.8} & \textbf{1.45} & \textbf{0.65} \\
\bottomrule
\end{tabular}}
\end{center}

\textsuperscript{a} QM9 56343 (0.38~eV) and 37992 (0.03~eV); without them, the other means change by at most 0.06~meV and 0.001 percentage points.

\textsuperscript{b} Trained and tested on VASP-computed densities of QM9 molecules\cite{QM9-VASP}, with splits of the same sizes; not directly comparable. BOA: 0.116\% on the densities used here, 0.134\% on QM9-VASP. eqDeepDFT: the equivariant version of DeepDFT.

\textsuperscript{c} Li et al.'s published QM9 model, re-evaluated with the one-step code used for our models (Methods).

\end{table}

One diagonalization brings the density closer to the self-consistent one, lowering the mean NMAE by a factor of 2.4--3.0 for all three models (Fig. 2a, dashed). The energy error is second order in the density error, so the differences between models widen (Fig. 2b). The median $\varepsilon_E$ is 0.52~meV for SR-ResNet, 0.14~meV for SwinUNETR-V2 and 0.036~meV for NeXtUNETR; the last two are 2.5--3 orders of magnitude below the 1~kcal mol$^{-1}$ (43~meV) threshold usually called chemical accuracy. NeXtUNETR's mean errors in the highest occupied and lowest unoccupied molecular orbital (HOMO and LUMO) energies and in the HOMO--LUMO gap are 2.8, 2.2 and 2.9~meV, about a quarter of SR-ResNet's; its median gap error is 1.8~meV.

\begin{figure}[!tbp]
{\centering
\makebox[\linewidth][c]{\includegraphics[width=155.9mm,height=0.55\textheight,keepaspectratio]{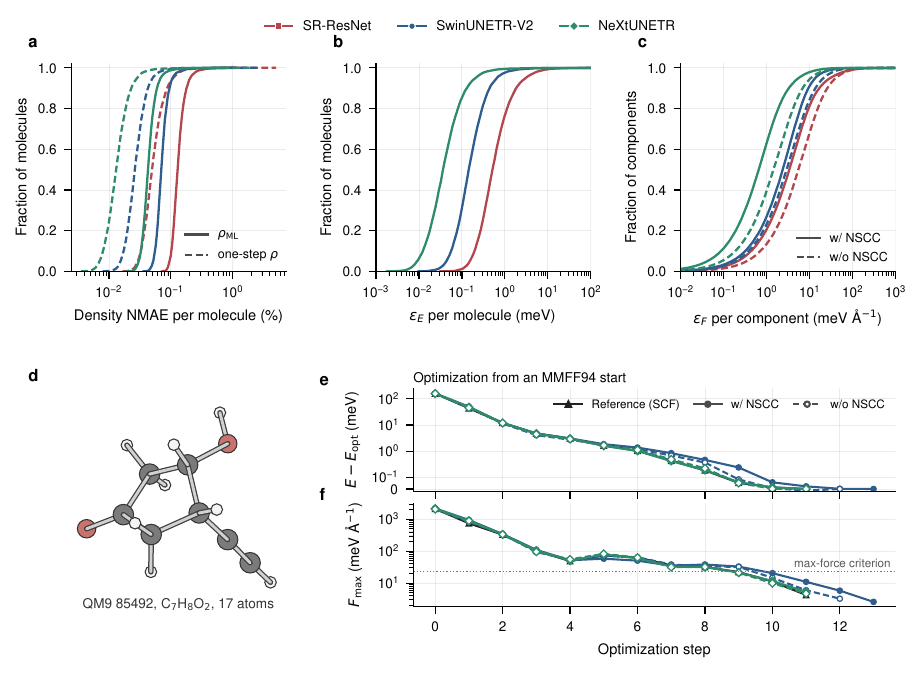}}\par}
\medskip\small
\textbf{Fig. 2 | Accuracy on QM9 and geometry optimization on the one-step surface.} \textbf{a}, Cumulative distribution of the density NMAE over the 10,000 QM9 test molecules; solid, predicted density $\rho_\mathrm{ML}$; dashed, density after the one-step diagonalization. \textbf{b}, Same for the one-step energy error $\varepsilon_E$. \textbf{c}, Same for the force error per Cartesian component; solid, with the non-self-consistency correction (w/ NSCC); dashed, without it (w/o NSCC). At most 4 molecules (\textbf{b}) and 110 components (\textbf{c}) per model lie beyond the axes, for SwinUNETR-V2 and NeXtUNETR almost all from the two molecules with a reference HOMO--LUMO gap below 1~eV. \textbf{d}, QM9 85492 (C$_{7}$H$_{8}$O$_{2}$), drawn at random from the molecules that every model optimizes to within 0.01~\AA{} of the reference optimum (four more in Supplementary Fig. 1). \textbf{e},\textbf{f}, Its optimization from an MMFF94 starting structure on the SCF and one-step surfaces, w/ and w/o NSCC: energy relative to each surface's optimum (\textbf{e}; symmetric logarithmic axis) and maximum force (\textbf{f}; dotted, the convergence criterion). Extended Data Table 2 summarizes all 200 molecules.

\end{figure}

The two molecules with reference HOMO--LUMO gaps below 1~eV, 0.38 and 0.03~eV, dominate the mean energy errors and the mean force errors with the NSCC. For them, one diagonalization can occupy the wrong orbitals, giving energy errors of 0.3--18~eV. Table 1 therefore also gives the means without them, and Extended Data Fig. 2 excludes them.

\subsection*{One-step forces}

With the NSCC, the median absolute errors of the force components are 3.37~meV~\AA{}$^{-1}$ for SR-ResNet, 2.37~meV~\AA{}$^{-1}$ for SwinUNETR-V2 and 0.65~meV~\AA{}$^{-1}$ for NeXtUNETR (Fig. 2c and Table 1). The median absolute reference component is 90~meV~\AA{}$^{-1}$. Without the NSCC, the medians of the same three models are 5.54, 2.93 and 1.45~meV~\AA{}$^{-1}$. The NSCC lowers the per-molecule mean absolute error of the forces for 88\% of the molecules with SR-ResNet, 86\% with SwinUNETR-V2 and 97\% with NeXtUNETR.

The mean over all molecules is nevertheless larger with the NSCC, because the correction amplifies the errors of the two small-gap molecules. Without them, the mean is smaller with the NSCC for every model (Table 1). The NSCC, which makes the force the exact gradient of the one-step energy, thus also brings it closer to the SCF force for most molecules.

\subsection*{Geometry optimization}

To test whether the one-step surfaces place their minima where the SCF surface does, we optimized 200 QM9 test molecules drawn at random (Methods). Each molecule was optimized on the SCF surface and on the one-step surfaces of both networks, w/ and w/o NSCC, and of the SAD guess as a control (Fig. 2d--f, Extended Data Table 2). All runs used geomeTRIC\cite{geomeTRIC} with the same criteria and started from an MMFF94 (ref.~\citenum{MMFF94}) relaxation of the reference optimum.

All optimizations with SwinUNETR-V2 and NeXtUNETR, w/ and w/o NSCC, converged in a median of 15.5--16 steps, as on the SCF surface (15.5). The final structures lie a median of 0.002--0.005~\AA{} from the optimum the reference reaches from the same start, and 97--98.5\% lie within 0.1~\AA{}, against 42.5\% of the starting structures. The runs shown also follow the reference step by step (Fig. 2e,f and Supplementary Fig. 1). QM9 contains only equilibrium structures, but every optimization starts from a distorted one; the agreement along the paths indicates that the predicted densities, and the energies and forces derived from them, remain accurate away from equilibrium.

Without a network, optimization fails: on the SAD surface the runs end, in the median, farther from the reference optimum than they started (w/ NSCC) or do not converge (w/o NSCC) (Extended Data Table 2). The one-step surfaces of the networks thus place their minima where the SCF surface does and are as easy to optimize on. This agreement comes from the accuracy of the predicted density, and near minima it does not depend on the NSCC.

\subsection*{Molecular dynamics}

Molecular dynamics probes the one-step surface farther from its minima. We ran 1~ps of constant-energy (NVE) dynamics of alanine dipeptide from four conformational basins, C7eq, C5, C7ax and ${\alpha}$L (Fig. 3), with forces from the SCF reference or from the one-step calculation with the NSCC. With 10 heavy atoms, the dipeptide has more than any QM9 molecule, and near room temperature it samples distorted structures unlike any in QM9.

\begin{figure}[!tbp]
{\centering
\makebox[\linewidth][c]{\includegraphics[width=180.0mm,height=0.55\textheight,keepaspectratio]{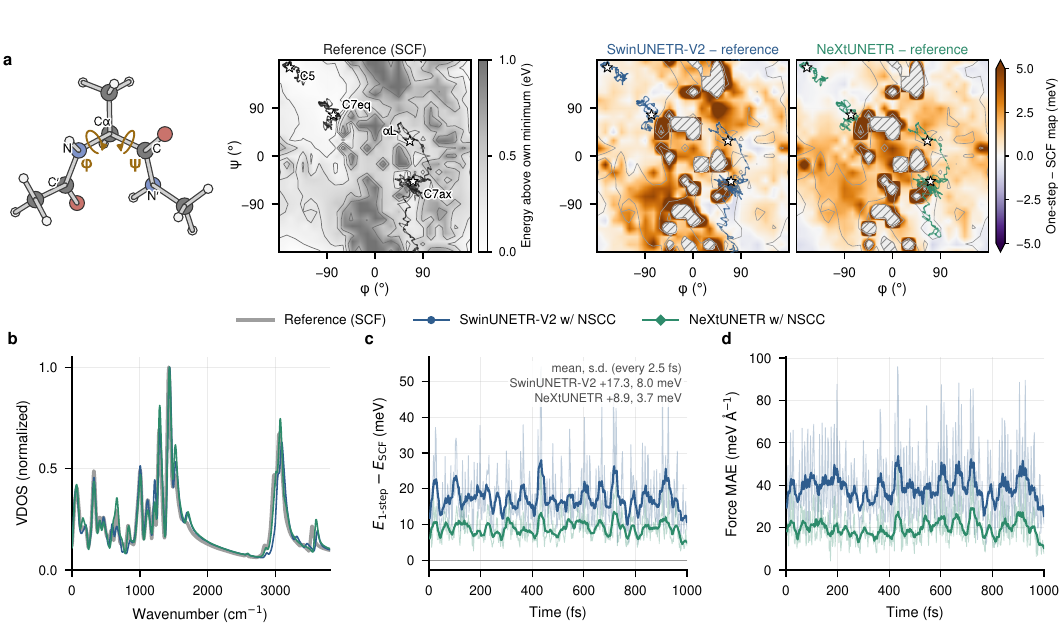}}\par}
\medskip\small
\textbf{Fig. 3 | Molecular dynamics with one-step forces.} 1~ps NVE trajectories of alanine dipeptide from four basins (C7eq, C5, C7ax, ${\alpha}$L), with forces from the SCF reference or from one Kohn--Sham step on the SwinUNETR-V2 or NeXtUNETR density, w/ NSCC; all runs from a basin share their initial velocities (Methods). \textbf{a}, Left: backbone dihedrals ${\varphi}$ (C$'$--N--C${\alpha}$--C) and ${\psi}$ (N--C${\alpha}$--C--N$'$). Middle: the reference trajectories on the SCF (${\varphi}$, ${\psi}$) energy map (contours at 0.15, 0.3, 0.5 and 0.75~eV); stars, starting structures. Right: error of each model's map (model $-$ SCF, each relative to its own minimum; color scale saturates at $\pm$5~meV) with that model's trajectories and the 0.3 and 0.75~eV SCF contours; hatched, more than 1~eV above the reference minimum. \textbf{b}, Vibrational density of states (VDOS) of the C7eq trajectories (reference in gray). \textbf{c},\textbf{d}, Error of the one-step energy (\textbf{c}) and forces (\textbf{d}; mean absolute error over force components) on the frames of the C7eq reference trajectory; light lines, every frame; dark lines, 25~fs running means. Extended Data Fig. 3 shows energy conservation and the dynamics w/o NSCC.

\end{figure}

The models reproduce the energy landscape and the motion on it. On the (${\varphi}$, ${\psi}$) map, the relative one-step energies deviate from the SCF ones by 1.5~meV for SwinUNETR-V2 and 1.3~meV for NeXtUNETR on average within 1~eV of the minimum (Fig. 3a). As in the reference, the trajectories from C7eq, C5 and C7ax stay in their basins, and those from ${\alpha}$L leave it in the same direction; individual trajectories are not expected to coincide, because nearby ones diverge. The vibrational density of states (VDOS), which reflects the curvature of the surface along the trajectory, has the same bands for both models as for the reference (Fig. 3b); the amide I and amide II band centroids lie within 9~cm$^{-1}$ of the reference in all four trajectories, below the 33~cm$^{-1}$ resolution of a 1~ps trajectory.

Along the reference C7eq trajectory, the one-step energies of SwinUNETR-V2 and NeXtUNETR are offset from the SCF energy by +17.3 and +8.9~meV on average (Fig. 3c), and their force errors, 38.7 and 19.7~meV~\AA{}$^{-1}$, do not grow with time (Fig. 3d). The errors exceed those on QM9, as expected for distorted structures, and NeXtUNETR's are about half of SwinUNETR-V2's in all four trajectories.

With the NSCC, the models conserve energy nearly as well as the reference: the standard deviation of the total energy is 0.06--0.10~meV atom$^{-1}$, against 0.05--0.07~meV atom$^{-1}$ (Extended Data Fig. 3). Without the NSCC, the deviation is 20--70 times larger, and the systems heat by 24--97~K. Energy conservation shows that the forces are consistent with the one-step energy, and the small energy errors show that this energy stays close to the SCF energy: one Kohn--Sham step can replace the SCF calculation in molecular dynamics of this molecule outside the training distribution.

\subsection*{Transfer beyond QM9}

We tested transfer on four test sets (Table 2), three from Li et al.\cite{Li2025} and one new: thermally distorted C$_{7}$O$_{2}$H$_{10}$ isomers from molecular dynamics; charged biomolecular fragments containing phosphorus and magnesium, both absent from QM9 (the MT set); water clusters of 30 molecules (W1500/30), three times larger than the largest QM9 molecule; and donor--bridge--acceptor (\mbox{D--B--A}) chains, in which an electron donor and an acceptor at the two ends polarize the density along a saturated or conjugated bridge. The test chains have longer bridges than any training chain, so this set tests extrapolation to longer molecules. Following Li et al., each model was applied directly (zero-shot), fine-tuned on a small target set and trained from scratch on that set.

\begin{table}[!tbp]
\small
\textbf{Table 2 | Transfer beyond QM9.} Density NMAE (\%) of the QM9 models applied directly (zero-shot) and after fine-tuning on the target training set, and of the same architectures trained from scratch on that set. One training run per fine-tuned or from-scratch entry (protocols in Methods). Best value per row in bold.

\begin{center}
\small
\resizebox{\ifdim\width>\linewidth\linewidth\else\width\fi}{!}{%
\begin{tabular}{llrrr}
\toprule
Test set & Training & SR-ResNet & SwinUNETR-V2 & NeXtUNETR \\
\midrule
C$_{7}$O$_{2}$H$_{10}$\textsuperscript{a} & zero-shot & \textbf{0.662} & 0.853 & 0.809 \\
 & fine-tuned & 0.244 & 0.228 & \textbf{0.167} \\
 & from scratch & 1.885 & \textbf{1.064} & 1.706 \\
MT\textsuperscript{b} & zero-shot & 2.168 & \textbf{1.518} & 3.501 \\
 & fine-tuned & 0.339 & 0.214 & \textbf{0.203} \\
 & from scratch & 3.478 & \textbf{0.993} & 1.931 \\
W1500/30\textsuperscript{c} & zero-shot & 0.723 & \textbf{0.432} & 0.637 \\
 & fine-tuned & 0.127 & \textbf{0.088} & 0.089 \\
 & from scratch & 1.283 & \textbf{0.403} & 0.981 \\
\mbox{D--B--A}\textsuperscript{d} & zero-shot & 0.299 & 0.356 & \textbf{0.250} \\
 & fine-tuned & 0.149 & 0.093 & \textbf{0.075} \\
 & from scratch & 0.725 & \textbf{0.540} & 0.860 \\
\bottomrule
\end{tabular}}
\end{center}

\textsuperscript{a} Ab initio molecular dynamics snapshots of 113 isomers; 4,520 test and 226 training structures.

\textsuperscript{b} Charged biomolecular fragments containing P and Mg; 110 test and 110 training structures.

\textsuperscript{c} The three 30-molecule clusters of W1500 (WDBase minima); trained on 3--13-molecule clusters (226 structures).

\textsuperscript{d} Donor--bridge--acceptor molecules; 144 test (C10, C12 bridges) and 353 training structures (C2--C8 bridges).

\end{table}

After fine-tuning, SwinUNETR-V2 and NeXtUNETR are more accurate than SR-ResNet on all four test sets, by up to a factor of 2.0 on the \mbox{D--B--A} chains (0.075 against 0.149\%). NeXtUNETR is the most accurate on C$_{7}$O$_{2}$H$_{10}$ and \mbox{D--B--A}, and the two are within 6\% of each other on MT and W1500/30. Trained from scratch, SwinUNETR-V2 is the most accurate model on every test set.

Without fine-tuning, the most accurate model depends on the test set: SwinUNETR-V2 on MT and W1500/30, NeXtUNETR on \mbox{D--B--A} and SR-ResNet on C$_{7}$O$_{2}$H$_{10}$. NeXtUNETR, the most accurate model on QM9, is the least accurate zero-shot on MT. Fine-tuning removes most of the differences between the two networks. Pretrained on QM9, SwinUNETR-V2 and NeXtUNETR adapt to new systems from 110--353 structures better than SR-ResNet and extrapolate to longer \mbox{D--B--A} bridges; trained from scratch on the same sets, every model is far less accurate.

\subsection*{Size transfer to water clusters}

To test zero-shot transfer in size, we built Wmin-PBE/3--100, a set of 53 water clusters of 3--100 molecules (Fig. 4). Up to 30 molecules, they are the putative global minima of the WDBase database\cite{WDBase,WDBase-web}; larger ones were grown from the 30-molecule minimum one water at a time with the TIP4P model\cite{TIP4P}, and all were relaxed with PBE (Methods). The largest has 300 atoms, ten times as many as the largest QM9 molecule. SwinUNETR-V2's NMAE is nearly independent of size, 0.50--0.59\% in each size band (3--10, 11--30, 31--50 and 60--100 molecules), whereas SR-ResNet's rises from 0.72 to 1.10\%. NeXtUNETR is the most accurate on the smallest clusters, 0.46\% for 3--10 molecules, but degrades beyond about 30 molecules, reaching 2.6\% for 60--100 molecules.

\begin{figure}[!tbp]
{\centering
\makebox[\linewidth][c]{\includegraphics[width=91.9mm,height=0.55\textheight,keepaspectratio]{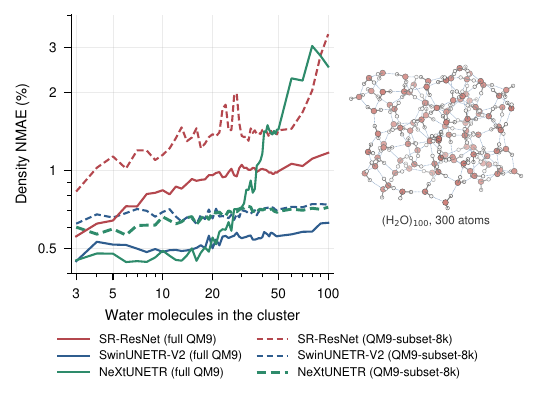}}\par}
\medskip\small
\textbf{Fig. 4 | Zero-shot transfer to water clusters far larger than any training molecule.} Density NMAE on Wmin-PBE/3--100 (log--log axes), 53 PBE-optimized water clusters: one of each size from 3 to 50 molecules and clusters of 60, 70, 80, 90 and 100 molecules, up to 300 atoms (right: the 100-molecule cluster, with hydrogen bonds dotted). None of the models was trained on water. Solid lines, models trained on the full QM9 training set; dashed lines, the same architectures trained on QM9-subset-8k (SwinUNETR-V2-292M, NeXtUNETR-67M and SR-ResNet-9M of Fig. 1).

\end{figure}

This degradation does not appear to be intrinsic to the architecture. Trained on QM9-subset-8k, NeXtUNETR and SwinUNETR-V2 both have nearly flat errors, 0.68 and 0.69\% over all clusters, whereas SR-ResNet degrades to 3.4\% at 100 molecules (Fig. 4, dashed). During NeXtUNETR's full-QM9 training, the NMAE on 60--100-molecule clusters fell to 0.68\% after 20 epochs and then rose to 2.6\% by the end of training, while that on small clusters fell to 0.46\% (Extended Data Fig. 4). As its accuracy on QM9 improves, NeXtUNETR thus appears to specialize to the molecular sizes of QM9.

\section*{Discussion}
Transformer-inspired hierarchical networks predict electron densities accurately enough that a single Kohn--Sham step reproduces the self-consistent electronic structure, from the total energy and forces to the orbital energies. The one-step forces recover the SCF minima in geometry optimization and drive energy-conserving molecular dynamics of a molecule with more heavy atoms than any training molecule. On QM9, NeXtUNETR outperforms SwinUNETR-V2 at lower cost.

Two features of the design matter. The Transformer-style hierarchy, which starts at half resolution and holds most parameters on coarse grids, keeps the cost per parameter low. The block design determines how well the parameters are used: in the same U-shaped network, with the same feed-forward layers, ConvNeXt blocks outperform Swin blocks that mix a 7 $\times$ 7 $\times$ 7 neighborhood by attention (NMAE 0.133 against 0.250\%). In computer vision, too, much of the advantage of Transformers comes from their macro-design and block recipe rather than from self-attention itself\cite{ConvNeXt,3DUX-Net,MedNeXt}. Attention may still help away from the training distribution, where SwinUNETR-V2 transfers best without fine-tuning to the charged fragments and the water clusters, but the size transfer of NeXtUNETR changes with its training (Extended Data Fig. 4), so single training runs cannot settle this.

The NSCC makes one-step forces usable for dynamics and lowers their error for most molecules. Minima are found equally well without it, but the uncorrected force is not the gradient of the one-step energy, and trajectories heat up. The NSCC costs one network backward pass per force evaluation.

A one-step evaluation costs far more than an MLIP evaluation, so this route complements MLIPs\cite{Unke2021,MACE-MP-0} rather than replacing them. In exchange, it keeps the Kohn--Sham Hamiltonian: orbital energies, the density and the properties derived from them come with the energy and forces, from one model trained only on densities. Unlike an SCF calculation, it has a fixed cost and is differentiable end to end. One-step surfaces could generate or check training data for MLIPs, or drive simulations that need electronic properties along the trajectory.

The study has limitations. Each model was trained once per condition, so differences of a few percent may lie within seed variation, and each molecular dynamics condition was run once, for 1~ps. One diagonalization can fail for near-degenerate frontier orbitals, as in two of the 10,000 QM9 molecules. All calculations used closed-shell PBE valence densities with pseudopotentials for molecules in periodic cells. Training on off-equilibrium structures and applying the approach to extended systems and other functionals are the next steps.

\section*{Methods}
\subsection*{Reference densities}

All densities are valence densities from Kohn--Sham density functional theory with the PBE functional\cite{PBE1996}, computed with PySCF\cite{PySCF2020} in the Gaussian-and-plane-wave (GPW) scheme\cite{GPW1997}: Goedecker--Teter--Hutter (GTH) pseudopotentials\cite{GTH1996} with PBE parameters\cite{Krack2005}, GTH-TZV2P basis sets\cite{Quickstep2005} and a plane-wave cutoff of 200~Ha (400~Ry) for the density. Valence electron counts are H 1, C 4, N 5, O 6, F 7, P 5 and Mg 10. Each molecule is placed with the centroid of its atoms at the center of an orthorhombic periodic cell whose edges are its extent along each axis plus 4~\AA{}, rounded up to an even number of grid intervals. The Hartree potential is evaluated by fast Fourier transform with the $\mathbf{G} = 0$ term omitted, and no Coulomb truncation is applied. The self-consistent field (SCF) is converged to 10$^{-11}$~Ha. The density is stored on the real-space grid of the calculation, whose spacing, ${\pi}$/20 bohr (0.083~\AA{}), resolves plane waves up to the 200~Ha cutoff.

QM9 (ref.~\citenum{QM9}) densities are those of Li et al.\cite{Li2025}, computed with this protocol on the B3LYP geometries of QM9, with their training, validation and test split of 123,835, 50 and 10,000 molecules (the split sizes of DeepDFT\cite{DeepDFT}). The C$_{7}$O$_{2}$H$_{10}$ and MT densities are also theirs; the MT labels use Fermi--Dirac smearing of 0.01~Ha and an SCF tolerance of 10$^{-7}$~Ha. We computed all other densities with PySCF 2.6.2 and a reimplementation of their generator, with the same settings, cell rule and grid; because PySCF 2.6.2 rounds the number of grid points per axis to an odd value, the grid size is fixed to the even value their version (PySCF 2.0.1) produces.

\subsection*{Input density}

The input to every network is a superposition of atomic densities (SAD)\cite{SAD-vanLenthe} sampled on the grid of the reference density, generated with PySCF 2.6.2 by a reimplementation of Li et al.'s procedure\cite{Li2025}. For each element, a free-atom density matrix is obtained in the minimal GTH-SZV basis from the one-electron Hamiltonian with the pseudopotential and spherically averaged occupations, and projected onto GTH-TZV2P. The atomic blocks are assembled into a block-diagonal density matrix, scaled to $N$ valence electrons, evaluated on the grid and renormalized to $N$ on the grid. For geometry optimization and molecular dynamics, the same density is built as a structure-factor sum, $\rho_\mathrm{SAD}(\mathbf{G}) = \sum_A \rho_{Z_A}(\mathbf{G})\,e^{-i\mathbf{G}\cdot\mathbf{R}_A}$, of element densities $\rho_Z$ precomputed in reciprocal space ($\mathbf{G}$ is a reciprocal-lattice vector and $Z_A$ the element of atom A), which makes it differentiable with respect to the nuclear positions $R_A$ by automatic differentiation.

\subsection*{Network architectures}

\textbf{Output.} Both networks predict a correction to the SAD density. The input $\rho_\mathrm{SAD}$ is centered in a zero-padded grid of the next multiple of 32 voxels (at least 64) along each axis, passed through the backbone $B$ and a 1 $\times$ 1 $\times$ 1 convolution head $h$, and added back: $\tilde\rho = \mathrm{ReLU}(\rho_\mathrm{SAD} + h(B(\rho_\mathrm{SAD})))$, cropped to the original grid. The output is rescaled to the number of valence electrons, $\rho_\mathrm{ML} = \tilde\rho\, N / (\Delta V \sum_\mathbf{r} \tilde\rho(\mathbf{r}))$, where $\Delta V$ is the voxel volume. The ReLU makes the density non-negative and the rescaling conserves the electron count exactly (Extended Data Fig. 1a).

\textbf{SwinUNETR-V2.} We used the MONAI\cite{MONAI} implementation of SwinUNETR-V2 (ref.~\citenum{SwinUNETR-V2}). A 2 $\times$ 2 $\times$ 2 patch embedding maps the input to feature size $C$ at half resolution; four encoder stages follow, each opened by a residual block of two 3 $\times$ 3 $\times$ 3 convolutions (the V2 addition) and built from Swin Transformer blocks\cite{Swin} that apply multi-head self-attention within 7 $\times$ 7 $\times$ 7-voxel windows, shifted by 3 voxels in alternate blocks. Patch merging after each stage halves the resolution and doubles the channels, giving $C\cdot\{1, 2, 4, 8, 16\}$ channels down to 1/32 of the input resolution. The stage depths are (5, 4, 2, 2), with (3, 6, 12, 24) attention heads; the depths were chosen in a preliminary search. As in MONAI's implementation of SwinUNETR\cite{SwinUNETR,MONAI}, the skip connections carry the input, the patch-embedded input and the outputs of the first three stages, and the stage-4 output feeds the bottleneck; all but the input are first layer-normalized without learned parameters. The input, the patch-embedded input and the outputs of the first two stages then pass through such residual blocks; the stage-3 output, at 1/16 resolution, enters the decoder without one, and the stage-4 output passes through one as the bottleneck. The decoder upsamples by transposed convolution, concatenates the skip features and applies a residual block, and a final 1 $\times$ 1 $\times$ 1 convolution maps its C output channels to the single channel passed to h. Where a residual block changes the channel count, its shortcut is a 1 $\times$ 1 $\times$ 1 convolution with group normalization. The convolutional blocks use group normalization\cite{GroupNorm} with $C/3$ groups instead of MONAI's default instance normalization.

\textbf{NeXtUNETR.} NeXtUNETR keeps the macro-design of SwinUNETR-V2 --- a 2 $\times$ 2 $\times$ 2 patchify stem to half resolution, four stages of depths (5, 4, 2, 2) with $C\cdot\{1, 2, 4, 8\}$ channels and a $16C$ bottleneck at 1/32 resolution --- and replaces each Swin block by a three-dimensional ConvNeXt block\cite{ConvNeXt}: a 7 $\times$ 7 $\times$ 7 depthwise convolution (zero-padded), group normalization ($C/3$ groups), a 1 $\times$ 1 $\times$ 1 convolution expanding the channels fourfold, GELU, a 1 $\times$ 1 $\times$ 1 projection and a residual connection, without layer scale or stochastic depth. The stem and the downsampling between stages are strided 2 $\times$ 2 $\times$ 2 convolutions followed by group normalization and GELU; the bottleneck is one ConvNeXt block. NeXtUNETR has none of the 3 $\times$ 3 $\times$ 3 residual blocks of SwinUNETR-V2: each stage begins directly with its ConvNeXt blocks, and the skip connections take each stage's output unchanged. Each decoder level applies a transposed convolution, concatenation with the skip features, a 1 $\times$ 1 $\times$ 1 fusion convolution and one ConvNeXt block, and a final transposed convolution returns to the input resolution before the 1 $\times$ 1 $\times$ 1 head.

\textbf{SR-ResNet.} The baseline is the super-resolution residual network of Li et al.\cite{Li2025}: 32 residual blocks of 32 channels (5 $\times$ 5 $\times$ 5 convolutions with circular padding) applied to the SAD density sampled at every second grid point, followed by twofold pixel-shuffle upsampling to the full grid. It does not add the input density to its output, and it rescales the output to the integral of its subsampled input (times eight) instead of to $N$; we evaluate it as published, without further rescaling. Table 1 uses Li et al.'s released QM9 model; for Fig. 1 the same architecture was retrained with their schedule (see QM9-subset-8k below).

\subsection*{Training}

The networks were implemented in PyTorch\cite{PyTorch} (version 2.3.1) with MONAI\cite{MONAI} (version 1.3.2). SwinUNETR-V2 and NeXtUNETR were trained with the same settings. The loss is the per-molecule normalized mean absolute error (NMAE) of the density on the unpadded grid (defined below, without the factor of 100). Training used AdamW\cite{AdamW} with learning rate 10$^{-4}$, weight decay 10$^{-5}$ and moment decay rates (${\beta}_{1}$, ${\beta}_{2}$) = (0.9, 0.999), batch size 1, for 50 epochs. The first epoch uses a constant learning rate of 10$^{-7}$ as a warm-up; the learning rate then follows a cosine decay from 10$^{-4}$ toward zero over the remaining 49 epochs, stepped once per epoch. Each training molecule is augmented on the fly, identically in input and label, by a rotation of the grid by multiples of 90$^\circ$ about each axis (drawn independently per axis) and, with probability 0.5, by an integer periodic shift of up to 4 voxels along each axis. During training the output is rescaled to the grid integral of the reference density, which equals $N$ up to the discretization, and the rescaling factor is capped at 100; at inference the factor is not capped (the target equals $N$ to within 10$^{-8}$). Of the end-of-epoch checkpoints, we kept the one with the lowest NMAE on the 50 validation molecules; for both full-QM9 models this is the last epoch.

\textbf{QM9-subset-8k.} The 8,000 molecules were drawn from the QM9 training split, stratified by size (five bins) and composition (molecules with F, and the CHNO, CHO, CHN and CH classes), with proportional allocation and at least five molecules per stratum (fixed random seed). The ten models of Fig. 1 and the ablation models of Extended Data Table 1 were trained on it for 50 epochs with the validation set above. SwinUNETR-V2 has widths $C = 24$, 48, 96 and 144 (18.5, 73.1, 291.7 and 655.9 million parameters) and NeXtUNETR $C = 48$, 96, 144 and 192 (17.5, 67.5, 150.1 and 265.4 million). SR-ResNet was trained with Li et al.'s schedule: Adam\cite{Adam} ((${\beta}_{1}$, ${\beta}_{2}$) = (0.9, 0.999), as in their released code) without weight decay, a first epoch at 10$^{-5}$ of the learning rate, then cosine decay over the remaining epochs, with their NMAE loss and augmentation. The learning rate was 0.1, the value Li et al. used for their QM9 variants (0.01 for their main model), and the best validation checkpoint was kept. SR-ResNet-70M has 64 channels and 64 residual blocks. The 3D U-Nets of Extended Data Table 1 have four levels, from the full grid down to 1/8 resolution, with 96, 192, 384 and 768 channels; each level applies two 3 $\times$ 3 $\times$ 3 convolutions with circular padding and ReLU, downsampling is by max pooling and upsampling by transposed convolution, and the output is formed as for SwinUNETR-V2 and NeXtUNETR. The plain version has no normalization; the residual version adds a residual connection around each pair of convolutions and group normalization (16 groups). Both used the optimizer, learning rate and loss of SwinUNETR-V2 and NeXtUNETR. The plain ResNet applies the 32 residual blocks of SR-ResNet to the full grid, without subsampling or upsampling, and was trained like SR-ResNet. The 500 test molecules of Fig. 1 were drawn uniformly at random from the 10,000 test molecules (fixed seed).

\textbf{Full QM9.} The models of Table 1 and of all later Results sections were trained on all 123,835 training molecules at $C = 96$ (SwinUNETR-V2 291.7 and NeXtUNETR 67.5 million parameters; breakdown in Extended Data Fig. 1e).

\subsection*{Density error, cost and precision}

The density error is the normalized mean absolute error on the reference grid, in percent, $\mathrm{NMAE} = 100 \times \sum_\mathbf{r} |\rho_\mathrm{ML}(\mathbf{r}) - \rho_\mathrm{ref}(\mathbf{r})| / \sum_\mathbf{r} \rho_\mathrm{ref}(\mathbf{r})$. For the one-step density, $\rho_\mathrm{ML}$ is replaced by the density of the one-step density matrix on the same grid points.

All network inference reported in the paper is in fp32 with TF32 tensor-core arithmetic disabled. Networks were trained with PyTorch's default, which enables TF32 for cuDNN convolutions. Switching it on changes the predictions of SwinUNETR-V2 and NeXtUNETR by at most 0.01\% and their mean NMAE by at most 0.001 percentage points (Supplementary Table 1), and the mean NMAE of Li et al.'s published model on the 500 molecules of Fig. 1 from 0.135 to 0.136\%; the SR-ResNets trained on QM9-subset-8k are more accurate when evaluated in TF32, the precision they were trained in (0.238 and 0.191\% instead of 0.320 and 0.198\%; Supplementary Table 1).

For Fig. 1, every model was run on the same 500 QM9 test molecules on one NVIDIA H200 NVL GPU (PyTorch 2.3.1, CUDA 12.1) with batch size 1 and the input already on the GPU. The forward time is the network alone, without the construction of the SAD density or the Kohn--Sham step; each molecule was timed three times and the median kept, and the reported value is the mean over the 500 molecules. The inputs have 483 distinct grid shapes and cuDNN autotuning was disabled. Floating-point operations were counted with PyTorch's FlopCounterMode on each input (convolutions, including depthwise and transposed ones, matrix products and attention); the operations not counted (normalization, activations, softmax and residual additions) contribute little.

\subsection*{One-step Kohn--Sham calculation}

From the predicted density $\rho_\mathrm{ML}$, the Kohn--Sham Hamiltonian matrix $H[\rho_\mathrm{ML}] = T + V_\mathrm{nl} + V_\mathrm{loc} + V_\mathrm{H}[\rho_\mathrm{ML}] + V_\mathrm{xc}[\rho_\mathrm{ML}]$ is built with the same GPW integration as the reference: the Hartree potential by fast Fourier transform on the grid, the PBE potential through libxc\cite{Libxc} with density gradients from the same transform, the local pseudopotential on the grid and the kinetic and non-local pseudopotential terms analytically. The generalized eigenvalue problem $HC = SC\varepsilon$ is solved once and the lowest $N/2$ orbitals are doubly occupied, $D = 2\,C_\mathrm{occ} C_\mathrm{occ}^\mathsf{T}$. The eigenvalues are the one-step orbital energies. The one-step energy is the Kohn--Sham energy functional of this density matrix,

\begin{equation*} E_\text{1-step} = \mathrm{Tr}[D(T + V_\mathrm{nl})] + \int v_\mathrm{loc}\,\rho_D\,\mathrm{d}\mathbf{r} + E_\mathrm{H}[\rho_D] + E_\mathrm{xc}[\rho_D] + E_\mathrm{NN}, \end{equation*}

where $\rho_D$ is the density of $D$ and $E_\mathrm{NN}$ the nuclear repulsion, evaluated by Ewald summation. Because $D$ is an idempotent density matrix with the correct number of electrons, $E_\text{1-step} \ge E_\mathrm{SCF}$ ($E_\text{1-step} - E_\mathrm{SCF} > 0$ for all 10,000 QM9 test molecules and all three models).

Neither PySCF nor GPU4PySCF provides the one-step energy or its gradient. The one-step calculation and all forces described below are our own code, which calls the libraries' routines for the grid density, the potentials, the integrals and their nuclear derivatives. The QM9 energies, orbital energies and forces (Table 1, Fig. 2a--c) were computed on the CPU with the PySCF build distributed by Li et al. (PySCF 2.0.1 with libxc 5.1.7), with which their reference energies are reproduced exactly and their published SR-ResNet errors are recovered (NMAE 0.138\% against 0.14\%; $\varepsilon_E$ 2.84 against 3.0~meV); with the reference density as input, the one-step energy differs from the SCF energy by 0.001~meV on average. Geometry optimization and molecular dynamics used our GPU implementation built on GPU4PySCF\cite{GPU4PySCF-Wu,GPU4PySCF-Li} (version 1.8.0, with PySCF 2.14.0, CuPy 14.1.1 and CUDA 12). In this implementation, the network and the diagonalization stay on the 200~Ha grid of the training data. $E_\text{1-step}$, the force terms at fixed $D$ and the response amplitudes are evaluated on a finer grid of 400~Ha ($\sqrt{2}$ times as many points per axis, rounded up to an even number); the derivative of $H[\rho_\mathrm{ML}]$ and the vector--Jacobian product stay on the 200~Ha grid. The reference SCF runs entirely at 400~Ha with GPU4PySCF's multigrid Kohn--Sham solver\cite{GPU4PySCF-MG}. The finer grid reduces the egg-box artifact\cite{SIESTA} of a grid fixed in space by a factor of 13--16 in energy and 5--6 in force, and the force remains the exact gradient of the reported energy (Supplementary Note 1 and Supplementary Fig. 2). These checks used an earlier CPU counterpart of the GPU implementation, run with PySCF 2.6.2, that evaluated the matrix $G$ of the NSCC (defined below) on the 200~Ha grid; its later version, which evaluates $G$ on the 400~Ha grid as the GPU implementation does, gives the CPU timings of Supplementary Table 2.

Within each study, the reference and the models use the same implementation and grid. Given the same density matrix, the CPU (PySCF 2.0.1) and GPU implementations differ at 200~Ha by 0.2--3.4~meV in energy and 0.3--6.9~meV~\AA{}$^{-1}$ in force (component mean absolute error, MAE; three structures), mainly through the discretization of the density on the multigrid; at 400~Ha, SCF calculations of alanine dipeptide (C7eq and C5) with the two implementations agree within 0.41~meV in energy and 0.1~meV~\AA{}$^{-1}$ in force (component MAE).

\subsection*{Forces and the non-self-consistency correction}

The one-step energy depends on the nuclear positions $R$ through the basis functions, the pseudopotentials and the nuclear repulsion, and through the input density $\rho_\mathrm{ML}(R) = f_\theta(\rho_\mathrm{SAD}(R))$, which in turn determines $D$. The force without the non-self-consistency correction (NSCC), denoted w/o NSCC, is minus the standard Kohn--Sham gradient evaluated for the one-step solution: the Hellmann--Feynman, pseudopotential, kinetic and basis-function (Pulay) derivatives contracted with $D$, the energy-weighted overlap term $-\mathrm{Tr}(W\,\partial S/\partial R)$ with $W = 2\sum_{i\in\mathrm{occ}} \varepsilon_i c_i c_i^\mathsf{T}$, and the nuclear repulsion. It treats $D$ as if it were self-consistent, neglecting the response of $D$ to $R$, both explicit and through the input density.

The force with the NSCC (w/ NSCC) is the exact gradient $-\mathrm{d}E_\text{1-step}/\mathrm{d}R$. With $G = \partial E_\text{1-step}/\partial D = H[\rho_D]$, the Kohn--Sham Hamiltonian matrix of the output density, $\mathrm{d}E_\text{1-step}/\mathrm{d}R = \partial E_\text{1-step}/\partial R|_D + \mathrm{Tr}[G\,\mathrm{d}D/\mathrm{d}R]$. Because $H$ is built from $\rho_\mathrm{ML}$ and not from $D$, $\mathrm{d}D/\mathrm{d}R$ follows from uncoupled first-order perturbation theory,

\begin{equation*} \mathrm{Tr}\!\left[G\,\frac{\mathrm{d}D}{\mathrm{d}R}\right] = 4 \sum_{i\in\mathrm{occ},\,a\in\mathrm{virt}} G_{ai}\,\frac{H^R_{ai} - \varepsilon_i S^R_{ai}}{\varepsilon_i - \varepsilon_a} - 2 \sum_{i,j\in\mathrm{occ}} G_{ij} S^R_{ij}, \end{equation*}

where $H^R$ and $S^R$ are the total derivatives of $H[\rho_\mathrm{ML}]$ and of the overlap matrix $S$ with respect to $R$, in the basis of the orbitals $C$. The force w/o NSCC replaces this term by the overlap term $-2\sum_{i\in\mathrm{occ}} \varepsilon_i S^R_{ii}$, its value at self-consistency; the NSCC is the difference. It vanishes when the input density is self-consistent, since then $G = H[\rho_\mathrm{ML}]$, $G_{ai} = 0$ and $G_{ij} = \varepsilon_i\delta_{ij}$. The explicit part of $H^R$ (basis functions and pseudopotentials at fixed $\rho_\mathrm{ML}$) is evaluated analytically by contracting the amplitude matrix $P = C_\mathrm{virt}\, z\, C_\mathrm{occ}^\mathsf{T}$, $z_{ai} = 4G_{ai}/(\varepsilon_i - \varepsilon_a)$, with the integral derivatives. The implicit part, through $\rho_\mathrm{ML}$, is $\int w(\mathbf{r})\,\partial\rho_\mathrm{ML}(\mathbf{r})/\partial R\,\mathrm{d}\mathbf{r}$ with the cotangent field $w = v_\mathrm{H}[\rho_P] + f_\mathrm{xc}[\rho_\mathrm{ML}]\,\rho_P$ (with the corresponding gradient terms of the PBE kernel), where $\rho_P$ is the density of $P$. This integral is obtained by a vector--Jacobian product (one backward pass) through the rescaling and the network, followed by the derivative of the SAD density.

For the SAD control of Extended Data Table 2, the same expressions are applied with $\rho_\mathrm{ML}$ replaced by $\rho_\mathrm{SAD}$. For SR-ResNet the vector--Jacobian product follows its own input (every second grid point) and output (no rescaling to $N$) conventions.

Every term was checked against finite differences of $E_\text{1-step}$. End to end, analytic and central-difference forces (step 0.002~\AA{}) agree within 0.03--1.8~meV~\AA{}$^{-1}$ (largest component deviation, at most 0.4\% of the component) on QM9 molecules and within 0.36~meV~\AA{}$^{-1}$ (component MAE) for alanine dipeptide; for the SAD control (step 0.001~\AA{}) their ratio is 0.9995--1.0001. The QM9 reference forces are the gradient w/o NSCC of the one-step calculation with the reference density as input, which agrees with PySCF's own SCF gradient in Li et al.'s build within 0.006--0.2~meV~\AA{}$^{-1}$ (largest component deviation) on five molecules. In the GPU implementation, the reference force is likewise our gradient (the w/o NSCC terms, exact at self-consistency) applied to the converged SCF density matrix, so that reference and model forces pass through the same code; it agrees with GPU4PySCF's own SCF gradient within 0.0003~meV~\AA{}$^{-1}$ (component MAE). Evaluating the SR-ResNet and SwinUNETR-V2 force errors on 30 QM9 molecules at 400 instead of 200~Ha changes the per-molecule force MAEs by at most 0.16~meV~\AA{}$^{-1}$, so the reported errors reflect the predicted density, not the integration grid. QM9 forces are reported as computed, without removing the net force.

\subsection*{Geometry optimization}

The 200 QM9 test molecules are the first, in a fixed random order, for which RDKit\cite{RDKit} assigned the bonds, an MMFF94 (ref.~\citenum{MMFF94}) starting structure could be built, the cell converged (below) and no heavy-atom dihedral differed by more than 60$^\circ$ between the starting structure and the reference optimum.

For each molecule, the reference optimum $x_\mathrm{ref}$ is the SCF minimum in a cell built by the rule of ``Reference densities'', obtained from the QM9 geometry by alternating SCF relaxation (with the optimizer and criteria below) and cell rebuilding until the cell no longer changed (at most three passes), and the starting structure is an MMFF94 relaxation of $x_\mathrm{ref}$ with RDKit, Kabsch-aligned\cite{Kabsch} to it. From this start, the structure was optimized on each surface of Extended Data Table 2 with geomeTRIC\cite{geomeTRIC} (version 1.1.1) in translation--rotation internal coordinates, with its default trust radius and convergence criteria (maximum force 23.1~meV~\AA{}$^{-1}$, RMS force 15.4~meV~\AA{}$^{-1}$, energy change 10$^{-6}$~Ha, maximum and RMS displacement 1.8 $\times$ 10$^{-3}$ and 1.2 $\times$ 10$^{-3}$~\AA{}), and at most 300 optimization steps. Net translation and rotation were projected out of every force. All runs used the 400~Ha GPU implementation; the reference SCF was converged to 10$^{-11}$~Ha from an atomic guess at every step, and the networks ran in fp32. Distances are all-atom Kabsch root-mean-square deviations (RMSDs) from the optimum the reference run reached from the same start (median 0.116~\AA{} for the starting structures).

On each network surface, 7--14 molecules ended more than 0.05~\AA{} from the reference optimum. For the 11 such molecules with SwinUNETR-V2 w/ NSCC, restarting each optimization from the other surface's end point showed that the two runs ended in different minima along soft torsions, which in all but one case exist on both surfaces.

\subsection*{Molecular dynamics}

Alanine dipeptide (Ace--Ala--NMe, 22 atoms) was started from its four gas-phase basins C7eq, C5, C7ax and ${\alpha}$L; each starting structure was built with MMFF94, the backbone dihedrals restrained to literature values, relaxed again without the restraint and then relaxed on the PBE/GTH surface in the cell below (200~Ha grid) to a maximum force of 20~meV~\AA{}$^{-1}$, which changed ${\varphi}$ and ${\psi}$ by at most 3.3$^\circ$. The cell is cubic, 13.30~\AA{} (160$^{3}$ grid), large enough for the molecule to rotate.

Trajectories were integrated with velocity Verlet\cite{Swope1982} in the NVE ensemble for 2,000 steps of 0.5~fs (1~ps). Initial velocities were drawn from a Maxwell--Boltzmann distribution at 600~K (one random seed shared by all methods for a basin) with the net momentum removed; by equipartition the trajectories settle at 258--284~K (except the runs w/o NSCC, which heat up). Forces came from the reference SCF (400~Ha, 10$^{-11}$~Ha, started from the previous step's density matrix) or from the one-step calculation on the SwinUNETR-V2 or NeXtUNETR density, with or without the NSCC. Because the grid is fixed in space, the forces do not sum exactly to zero; rather than projecting out the net force, which would break consistency with the reported energy, we recentered the centroid in the cell and removed the center-of-mass velocity at every step.

Energy conservation is measured by the standard deviation of the total energy per atom over the second half of each trajectory (not detrended). The vibrational density of states is the Fourier transform of the windowed, mass-weighted velocity autocorrelation function (velocities every 2.5~fs; resolution 33~cm$^{-1}$); band positions are intensity-weighted centroids over fixed windows (amide II, 1,250--1,550~cm$^{-1}$; amide I, 1,550--1,800~cm$^{-1}$). The (${\varphi}$, ${\psi}$) maps were computed on a 15$^\circ$ grid (576 points) of MMFF94 structures relaxed with the two dihedrals constrained, with the SCF and each model's one-step energy (with the implementation and cell of the dynamics), each relative to its own minimum; mean absolute map errors are taken over the 541 points within 1~eV of the SCF minimum. The per-frame errors of Fig. 3c,d are the one-step energies and forces (w/ NSCC) evaluated on the 400 frames (every 2.5~fs) of each reference trajectory.

\subsection*{Single-point timing}

Single-point wall times (Supplementary Table 2) were measured on the 200 QM9 molecules of the geometry-optimization set, each at its reference-SCF minimum, on one NVIDIA A40 (48~GB), A100 SXM4 (80~GB) or H100 NVL GPU and on 16 cores of an AMD EPYC 7543 CPU. On each device the reference SCF and the one-step calculation used the same implementation: on the GPUs that of the geometry optimizations, on the CPU the PySCF 2.6.2 implementation above. Both used the settings of the geometry optimizations (SCF converged to 10$^{-11}$~Ha from the atomic guess on the 400~Ha grid). Each time is that of a new molecule in a running program; program start, model loading, file input and cell construction are excluded, and performance options that leave the results unchanged (forces within 0.02~meV~\AA{}$^{-1}$) were used. Every calculation was run once per molecule, after a warm-up on a molecule outside the set and with per-molecule caches emptied before each calculation, and the table reports the median over the 200 molecules. Network forward and backward passes are also reported separately.

\subsection*{Transfer benchmarks}

\textbf{Datasets.} \emph{C$_{7}$O$_{2}$H$_{10}$}\cite{Li2025}: 113 isomers $\times$ 50 snapshots from ab initio molecular dynamics; following Li et al., the first two snapshots of each trajectory are used for training (226), the third for validation (113) and the last 40 for testing (4,520). \emph{MT}\cite{Li2025}: charged biomolecular fragments from a GTP-bound microtubule (total charge $-$4) containing C, H, N, O, P and Mg; the first 10 structures of each of Li et al.'s runs are used for training (110) and the 11th for validation, and the test set is 10 structures per run drawn from their test range in ten index-quantile bins (110). \emph{W1500}: water clusters from the WDBase database of water-cluster minima\cite{WDBase,WDBase-web} (its 1~kcal mol$^{-1}$ files), taking the first min(count, $\lfloor 1500/n \rfloor$) structures of each n-molecule file, which reproduces Li et al.'s counts; training on $n = 3$--13 (226), validation on $n = 14$ (17) and testing on the three $n = 30$ clusters (W1500/30). \emph{Wmin-PBE/3--100}: for $n = 3$--30 the putative global minima of the TTM2.1-F potential listed in WDBase (for $n = 26$--30 from an incomplete search), and for $n = 31$--50, 60, 70, 80, 90 and 100 clusters grown deterministically from the 30-molecule minimum, one rigid water at a time: among the TIP4P\cite{TIP4P} placements within 0.02~eV of the lowest-energy one, the one giving the smallest cell was kept, with rigid-body reorganization every five additions and a beam of three candidate structures. All Wmin-PBE clusters were relaxed on the PBE/GTH surface (200~Ha) to a maximum force of 50~meV~\AA{}$^{-1}$; W1500 uses the WDBase geometries as given. Both were labeled with the protocol above. \emph{\mbox{D--B--A} chains} (constructed here): six donors (NMe$_{2}$, NH$_{2}$, OMe, OH, Me, F) and six acceptors (NO$_{2}$, CN, CF$_{3}$, CHO, COOH, CONH$_{2}$) at the ends of saturated (alkane) and conjugated (\emph{trans}-polyene) bridges of C2, C4, \ldots{}, C12. The parent chains, built with RDKit (ETKDG version 3\cite{ETKDG} and MMFF94), were relaxed on the PBE/GTH surface (200~Ha) to a maximum force of 50~meV~\AA{}$^{-1}$; donor and acceptor substituents were relaxed with the bridge frozen (to the same threshold, except two NMe$_{2}$ fragments that stopped at 56 and 84~meV~\AA{}$^{-1}$ after 80 steps) and the doubly substituted molecules assembled from them. All molecules of a series (one bridge type and length) share one cell, centered on the bridge, whose edges are the largest extent in the series plus 4~\AA{}. Models were trained on all C2--C8 molecules (parent, singly and doubly substituted; 353 training, 39 validation drawn at random) and tested on the 144 doubly substituted molecules with C10 and C12 bridges.

\textbf{Protocols.} Zero-shot results apply the QM9 models of Table 1 directly. Fine-tuning starts from them and trains for 50 epochs on the target training set, with fresh optimizer and scheduler states. SwinUNETR-V2 and NeXtUNETR use the loss, optimizer, schedule and augmentation of pre-training (AdamW, learning rate 10$^{-4}$), and SR-ResNet Li et al.'s protocol (Adam, learning rate 0.01, a first epoch at 10$^{-5}$ of that rate, then cosine decay, with the Adam settings above). Models trained from scratch use the same settings from random initialization. For every model, the end-of-epoch checkpoint with the lowest validation NMAE is evaluated. Wmin-PBE/3--100 was evaluated zero-shot only. On the transfer sets, only the density was evaluated.

\subsection*{Statistics and reproducibility}

Each model and condition was trained once, and each molecular dynamics condition was run once; no seed variation is reported. QM9 results include all 10,000 test molecules; the two molecules with a reference HOMO--LUMO gap below 1~eV are excluded only from the parenthesized means of Table 1 and from Extended Data Fig. 2, by that criterion alone.

\section*{Data availability}
The QM9, C$_{7}$O$_{2}$H$_{10}$ and MT reference densities are those of Li et al.\cite{Li2025}, deposited in CaltechDATA\cite{Li-QM9-data,Li-C7-data,Li-MT-data}; their released SR-ResNet models are on Figshare\cite{Li-models}. The densities computed here (the W1500 clusters used, Wmin-PBE/3--100 and the \mbox{D--B--A} chains), the structures and the source data will be made publicly available upon publication.

\section*{Code availability}
The model definitions, the training code, the one-step and force code and the analysis scripts will be made publicly available upon publication.

\section*{Acknowledgements}
This work was supported by a KIAS Individual Grant (AP110201) via the Center for Artificial Intelligence and Natural Sciences at Korea Institute for Advanced Study.

\section*{Author contributions}
J.W. conceived the study, carried out all calculations and analyses, and wrote the manuscript.

\section*{Competing interests}
The author declares no competing interests.

\clearpage
\section*{Extended Data}

\begin{center}
\makebox[\linewidth][c]{\includegraphics[width=180.0mm,height=0.62\textheight,keepaspectratio]{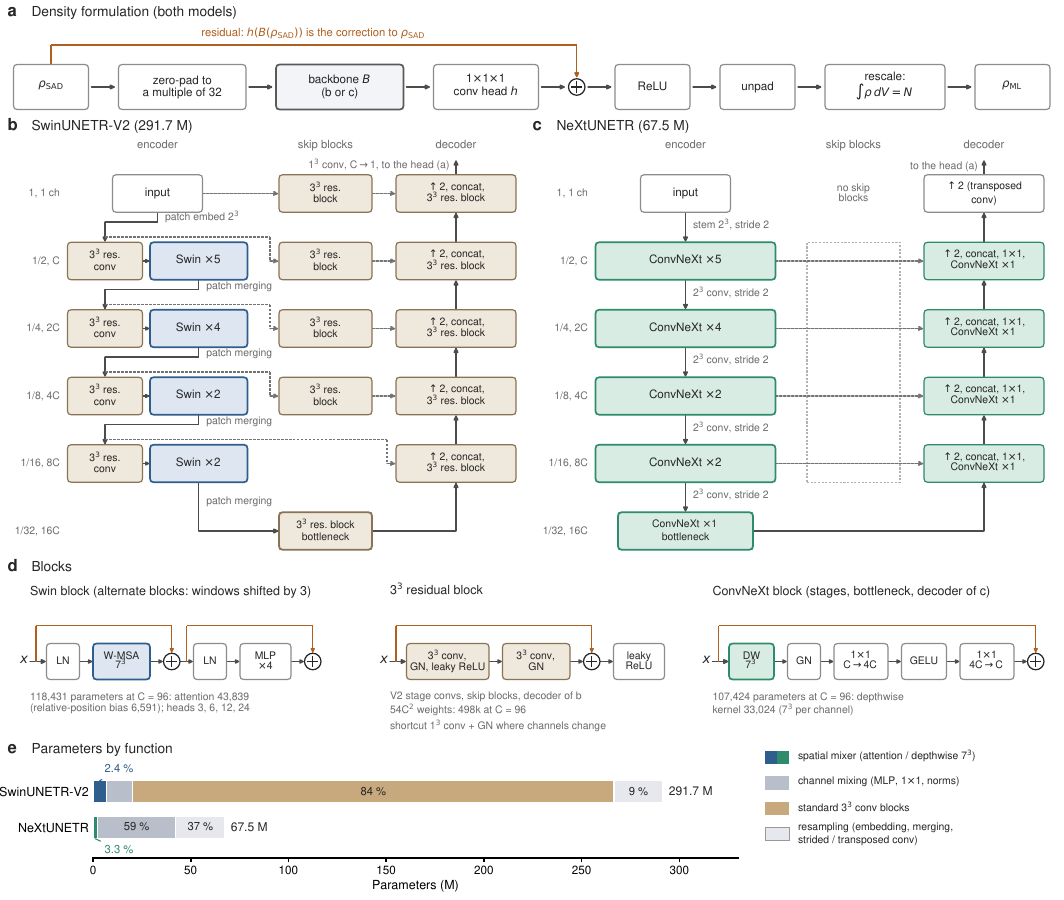}}
\end{center}
\small
\textbf{Extended Data Fig. 1 | SwinUNETR-V2 and NeXtUNETR.} $k^3$ denotes a $k \times k \times k$ kernel or patch. \textbf{a}, Density formulation shared by both models: the zero-padded SAD density is refined by the backbone $B$ and a 1$^{3}$ convolution head $h$ into a correction that is added back, made non-negative, cropped and rescaled to the number of valence electrons $N$ (Methods). \textbf{b},\textbf{c}, The backbones of SwinUNETR-V2 (\textbf{b}) and NeXtUNETR (\textbf{c}) on a common grid: rows are resolutions from the input grid to 1/32 (channel counts in multiples of $C = 96$); columns are the encoder, the blocks on the skip connections and the decoder. SwinUNETR-V2 places 3$^{3}$ residual blocks on the skips (except at 1/16), in the bottleneck and in the decoder, layer-normalizes the features leaving the encoder (without learned parameters) and ends in a 1$^{3}$ convolution from C channels to one; NeXtUNETR uses ConvNeXt blocks in the bottleneck and decoder and passes the skips unchanged. \textbf{d}, Swin block (alternate blocks shift the windows by 3 voxels), 3$^{3}$ residual block (with a 1$^{3}$ convolution on the shortcut where the channel count changes) and ConvNeXt block, with parameter counts at $C = 96$ for the first stage. \textbf{e}, Parameters by function. Spatial mixing takes 2.4\% (attention) and 3.3\% (depthwise 7$^{3}$ kernels) of the parameters of SwinUNETR-V2 and NeXtUNETR, respectively; SwinUNETR-V2's size comes from its standard (non-depthwise) 3$^{3}$ convolutions (84\%), with 44\% in the bottleneck block alone.

\normalsize

\clearpage
\begin{center}
\makebox[\linewidth][c]{\includegraphics[width=163.8mm,height=0.62\textheight,keepaspectratio]{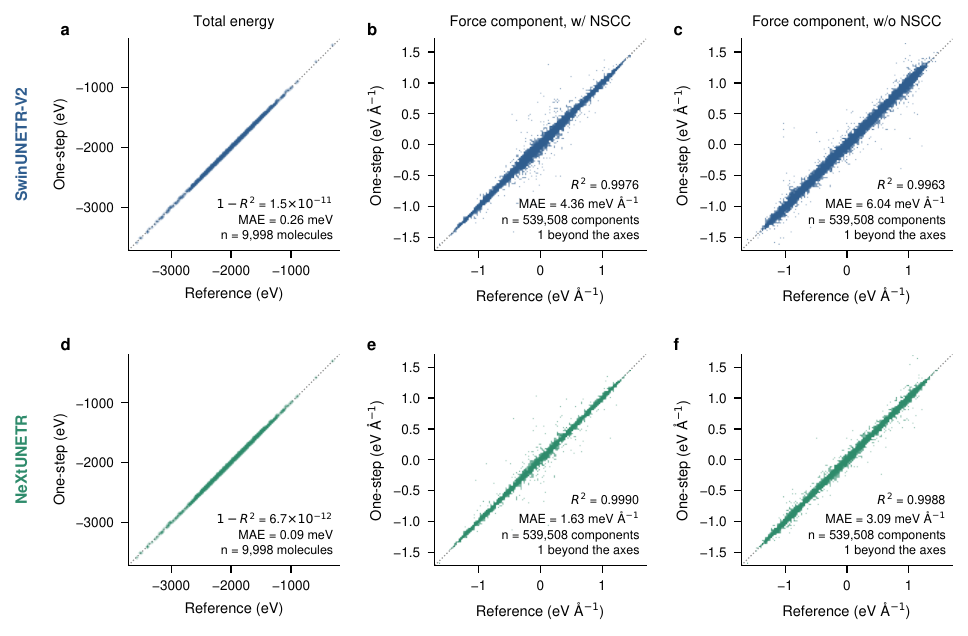}}
\end{center}
\small
\textbf{Extended Data Fig. 2 | One-step energies and forces against the reference SCF, QM9 test set.} Total energy per molecule (\textbf{a},\textbf{d}) and Cartesian force components with (\textbf{b},\textbf{e}) and without (\textbf{c},\textbf{f}) the non-self-consistency correction, for SwinUNETR-V2 (top) and NeXtUNETR (bottom). The two molecules with a reference HOMO--LUMO gap below 1~eV (QM9 56343 and 37992; energy errors 0.6--5.3~eV with these two models and 0.3--18~eV including SR-ResNet, force-component errors up to 300~eV \AA{}$^{-1}$ w/ NSCC) are excluded; with them, the force-component R$^{2}$ w/ NSCC becomes negative. R$^{2}$ (1 $-$ R$^{2}$ for energies) and MAE are over the remaining 9,998 molecules (539,508 components); Table 1 reports all 10,000. Dotted: $y = x$.

\normalsize

\clearpage
\begin{center}
\makebox[\linewidth][c]{\includegraphics[width=180.0mm,height=0.62\textheight,keepaspectratio]{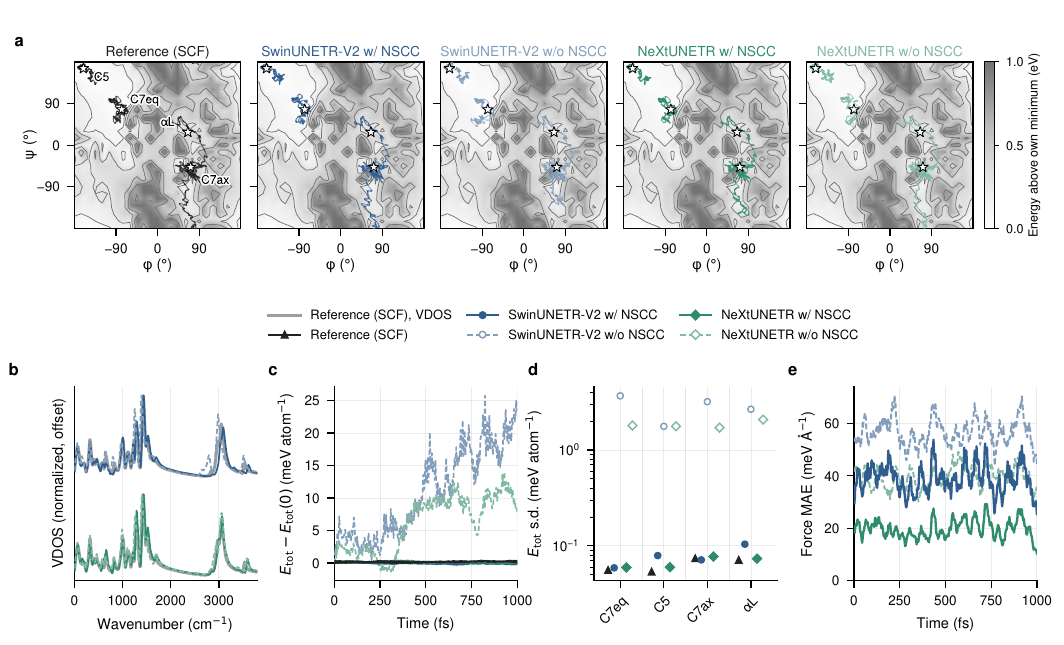}}
\end{center}
\small
\textbf{Extended Data Fig. 3 | Energy conservation requires the non-self-consistency correction.} \textbf{a}, Backbone dihedrals of the four alanine-dipeptide trajectories of each method on its own (${\varphi}$, ${\psi}$) map: the SCF map for the reference, the model's one-step map for the model (w/ and w/o NSCC share it). \textbf{b}, Vibrational density of states (VDOS) of the C7eq trajectories w/ and w/o NSCC (SwinUNETR-V2 above, NeXtUNETR below, offset vertically), each over the reference (gray). \textbf{c}, Total energy per atom relative to its initial value along the C7eq trajectories. \textbf{d}, Standard deviation of the total energy over the second half of each trajectory, all four trajectories. \textbf{e}, Force error (component MAE, 25~fs running mean) on the frames of the C7eq reference trajectory, w/ and w/o NSCC. Without the NSCC, the s.d. is 1.7--3.7~meV atom$^{-1}$ (reference 0.05--0.07~meV atom$^{-1}$) and the systems heat by 24--97~K (not shown). Settings as in Fig. 3; one trajectory per condition.

\normalsize

\clearpage
\begin{center}
\makebox[\linewidth][c]{\includegraphics[width=157.2mm,height=0.62\textheight,keepaspectratio]{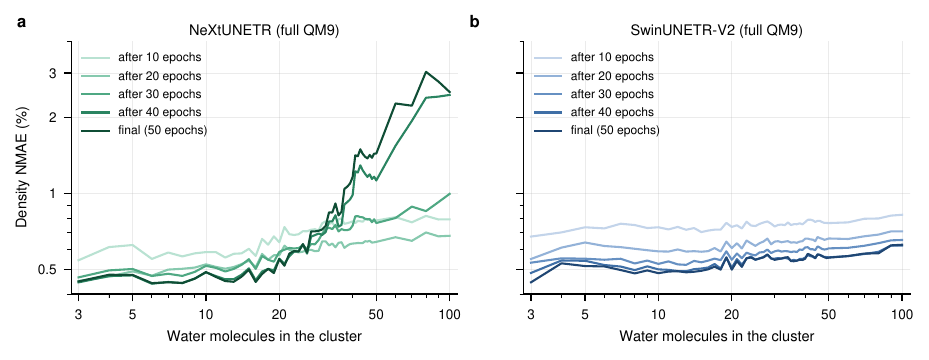}}
\end{center}
\small
\textbf{Extended Data Fig. 4 | Size extrapolation during training.} Density NMAE on Wmin-PBE/3--100 for full-QM9-trained NeXtUNETR (\textbf{a}) and SwinUNETR-V2 (\textbf{b}) after 10, 20, 30 and 40 epochs and at the end of training (50 epochs). The NeXtUNETR error on small clusters falls with training (3--10 molecules: from 0.58\% to 0.46\%), whereas that on large clusters is lowest after 20 epochs and then rises (60--100 molecules: from 0.68\% to 2.57\%). The SwinUNETR-V2 error falls in every size range up to 40 epochs and then levels off (60--100 molecules: from 0.80\% to 0.59\%).

\normalsize

\clearpage
\textbf{Extended Data Table 1 | Structural ablations on QM9-subset-8k.} Density NMAE (\%) on the 500 QM9 test molecules of Fig. 1. Each variant is compared with the first row of its block: relative change of the mean (${\Delta}$ mean) and share of molecules on which the variant is worse. One training run each (50 epochs); fp32 evaluation.

\begin{center}
\footnotesize
\setlength{\tabcolsep}{4pt}
\resizebox{\ifdim\width>\linewidth\linewidth\else\width\fi}{!}{%
\begin{tabular}{lp{0.42\linewidth}rrrr}
\toprule
Family & Model & Params (M) & NMAE mean / median (\%) & ${\Delta}$ mean & Worse on \\
\midrule
SwinUNETR-V2 & reference (SwinUNETR-V2-292M of Fig. 1) & 291.7 & 0.204 / 0.190 & --- & --- \\
 & without the 3$^{3}$ residual blocks on the downsampled skips and in the bottleneck\textsuperscript{a} & 153.8 & 0.209 / 0.196 & +2\% & 69\% \\
NeXtUNETR & reference (ConvNeXt encoder blocks) & 67.5 & 0.133 / 0.124 & --- & --- \\
 & encoder blocks replaced by Swin blocks\textsuperscript{b} & 73.2 & 0.250 / 0.236 & +88\% & 99.8\% \\
SR-ResNet & reference (half-resolution input, $\times$2 pixel-shuffle upsampling)\textsuperscript{c} & 9.4 & 0.320 / 0.309 & --- & --- \\
 & plain ResNet on the full grid (no upsampling) & 8.3 & 0.301 / 0.276 & $-$6\% & 29\% \\
3D U-Net & reference (plain 3D U-Net, base width 96) & 50.4 & 0.499 / 0.466 & --- & --- \\
 & 3D residual U-Net with group normalization (base width 96) & 51.2 & 0.259 / 0.243 & $-$48\% & 0\% \\
\bottomrule
\end{tabular}}
\end{center}

\textsuperscript{a} Residual blocks on the 1/2, 1/4 and 1/8 skip connections and the bottleneck block (138 M parameters, 127 M of them in the bottleneck block).

\textsuperscript{b} Windowed attention in a 7 $\times$ 7 $\times$ 7 window; everything outside the encoder blocks unchanged.

\textsuperscript{c} Li et al.'s architecture (32 channels, 32 residual blocks) with their schedule; SR-ResNet-9M of Fig. 1.

\clearpage
\textbf{Extended Data Table 2 | Geometry optimization on one-step surfaces.} 200 QM9 test molecules optimized with geomeTRIC from MMFF94 starting structures. RMSD: Kabsch root-mean-square deviation from the optimum the reference reaches from the same start. SAD: the one-step calculation on the SAD density, without a network. Converged: all criteria met within 300 steps. Steps include the starting structure; 301 means not converged within 300 steps. w/ and w/o NSCC: with and without the non-self-consistency correction. p90: 90th percentile.

\begin{center}
\footnotesize
\setlength{\tabcolsep}{4pt}
\resizebox{\ifdim\width>\linewidth\linewidth\else\width\fi}{!}{%
\begin{tabular}{lrrrrr}
\toprule
Surface & Converged & Steps, median (max) & RMSD, median / p90 (\AA{}) & Within 0.1~\AA{} & Heavy-atom RMSD, median (\AA{}) \\
\midrule
MMFF94 start & --- & --- & 0.116 / 0.335 & 42.5\% & 0.078 \\
Reference SCF & 200/200 & 15.5 (61) & 0 (by definition) & --- & --- \\
SwinUNETR-V2 w/ NSCC & 200/200 & 16 (45) & 0.003 / 0.034 & 98.0\% & 0.002 \\
SwinUNETR-V2 w/o NSCC & 200/200 & 16 (63) & 0.005 / 0.038 & 97.0\% & 0.003 \\
NeXtUNETR w/ NSCC & 200/200 & 15.5 (52) & 0.002 / 0.021 & 98.5\% & 0.001 \\
NeXtUNETR w/o NSCC & 200/200 & 15.5 (53) & 0.003 / 0.032 & 98.0\% & 0.002 \\
SAD w/ NSCC & 199/200 & 22 (301) & 0.156 / 0.472 & 23.0\% & 0.120 \\
SAD w/o NSCC & 5/200 & 301 (301) & 0.245 / 0.416 & 2.5\% & 0.181 \\
\bottomrule
\end{tabular}}
\end{center}

\clearpage
\section*{Supplementary Information}
\subsection*{Supplementary Note 1: the 400 Ha evaluation grid}
\emph{Why.} The labels, the SAD input and the network outputs are all defined on the uniform 200~Ha (400~Ry) grid of the reference densities of Li et al.\cite{Li2025}. A grid fixed in space is not translation invariant: moving a molecule rigidly relative to the grid points changes the multigrid energy and forces, with the period of the grid (the egg-box effect). We encountered it in geometry optimization on the 200~Ha grid, where the maximum force stalled at 12--28~meV~\AA{}$^{-1}$ while the energy kept decreasing.

\emph{Size of the effect.} Supplementary Fig. 2 quantifies it for the relaxed alanine dipeptide (C7eq), translated rigidly by one grid spacing along each axis (0.144~\AA{} along the cell diagonal, 13 points). A rigid translation changes no physical quantity, so every variation is an artifact. These measurements used our CPU implementation (PySCF 2.6.2).

\begin{itemize}
\item At 200~Ha, the energy varies by 21.5~meV peak to peak (reference SCF) and 21.3~meV (one-step, SwinUNETR-V2), and returns to within 0.02~meV after one period.
\item Per-atom forces (net force removed) vary by 26.8 and 27.3~meV~\AA{}$^{-1}$ RMS.
\item The variation is concentrated on oxygen (up to 200~meV~\AA{}$^{-1}$) and nitrogen (90~meV~\AA{}$^{-1}$), whose GTH pseudopotential cores are the hardest and whose valence densities are therefore the sharpest relative to the grid spacing. Carbon and hydrogen vary by 22--30~meV~\AA{}$^{-1}$.
\end{itemize}

\emph{What the 400~Ha grid changes.} On a 400~Ha grid ($\sqrt{2}$ times as many points per axis), the same translation changes the energy by 1.3~meV (SCF) and 1.6~meV (one-step) and per-atom forces by 4.4 and 5.4~meV~\AA{}$^{-1}$ RMS (largest 38 and 45~meV~\AA{}$^{-1}$, on oxygen). The artifact thus falls 13--16-fold in energy and 5--6-fold in force.

The artifact is common to the reference and the one-step calculation, so it cancels in the comparisons the paper reports:

\begin{itemize}
\item the electronic-energy error of the one-step calculation (SwinUNETR-V2, alanine dipeptide) is 2.97~meV on average at 200~Ha and 2.96~meV at 400~Ha, varying by 0.42~meV over the translation on both grids;
\item on 30 QM9 molecules the per-molecule force MAE against the reference changes by at most 0.16~meV~\AA{}$^{-1}$ between the two grids.
\end{itemize}

The 400~Ha grid therefore matters for the energy surface itself, which geometry optimization and molecular dynamics follow, and not for the accuracy figures.

\emph{How.} The networks are not retrained: the SAD input, the prediction $\rho_\mathrm{ML}$, the Kohn--Sham Hamiltonian matrix $H[\rho_\mathrm{ML}]$ and its diagonalization stay on the 200~Ha grid, where the training data are defined, and only the evaluation of the resulting density matrix $D$ moves to the finer grid.

\begin{itemize}
\item \emph{Energy.} $E_\text{1-step} = E[D]$ is assembled on the 400~Ha grid. The kinetic and nonlocal-pseudopotential terms are analytic and do not depend on the grid. The local pseudopotential, Hartree and exchange--correlation terms use the density of $D$ collocated on GPU4PySCF's multigrid at 400~Ha.
\item \emph{Grid size.} The fine mesh is the label mesh scaled by $\sqrt{2}$ per axis and rounded up to an even number of points, since odd meshes break the parity on which the label grid's density representation relies.
\item \emph{Force.} The force is the exact gradient of this energy.
\begin{itemize}
\item The w/o NSCC term (fixed $D$) is evaluated on the 400~Ha grid.
\item The NSCC uses $G = \partial E_\text{1-step}/\partial D$, the Kohn--Sham Hamiltonian matrix of $D$ on the 400~Ha grid, in the response amplitudes of Methods.
\item The derivative of $H[\rho_\mathrm{ML}]$ with respect to the nuclear positions, and the vector--Jacobian product through the network and the SAD, are taken on the 200~Ha grid, where $\rho_\mathrm{ML}$ is defined.
\end{itemize}
\item \emph{Reference.} The reference SCF runs entirely at 400~Ha. Its force is our gradient at the converged density matrix, which agrees with GPU4PySCF's own SCF gradient within 0.0003~meV~\AA{}$^{-1}$ (component MAE).
\end{itemize}

\emph{Validation and cost.}

\begin{itemize}
\item For alanine dipeptide, the analytic force agrees with central finite differences of the 400~Ha energy within 0.36~meV~\AA{}$^{-1}$ (component MAE, SwinUNETR-V2 w/ NSCC, h = 0.002~\AA{}). This is the same residual as that of the 200~Ha path against its own energy.
\item Building the response amplitudes from the 200~Ha Kohn--Sham Hamiltonian matrix instead would leave an inconsistency of 1.15~meV~\AA{}$^{-1}$.
\item Along the translation, the analytic net force reproduces the derivative of the energy within the error of the 13-point difference.
\item The finer evaluation adds 4\% to an energy-and-force step (alanine dipeptide, SwinUNETR-V2 w/ NSCC, GPU implementation on an A100: 6.72 against 6.44 s).
\item The RMS of what remains of the artifact (4.4--5.4~meV~\AA{}$^{-1}$; 1.3--1.6~meV in energy) is well below the geometry-optimization maximum-force criterion (23~meV~\AA{}$^{-1}$), although the largest per-atom deviation over a full grid period, on oxygen, is 38--45~meV~\AA{}$^{-1}$.
\end{itemize}

\clearpage
\begin{center}
\makebox[\linewidth][c]{\includegraphics[width=154.6mm,height=0.62\textheight,keepaspectratio]{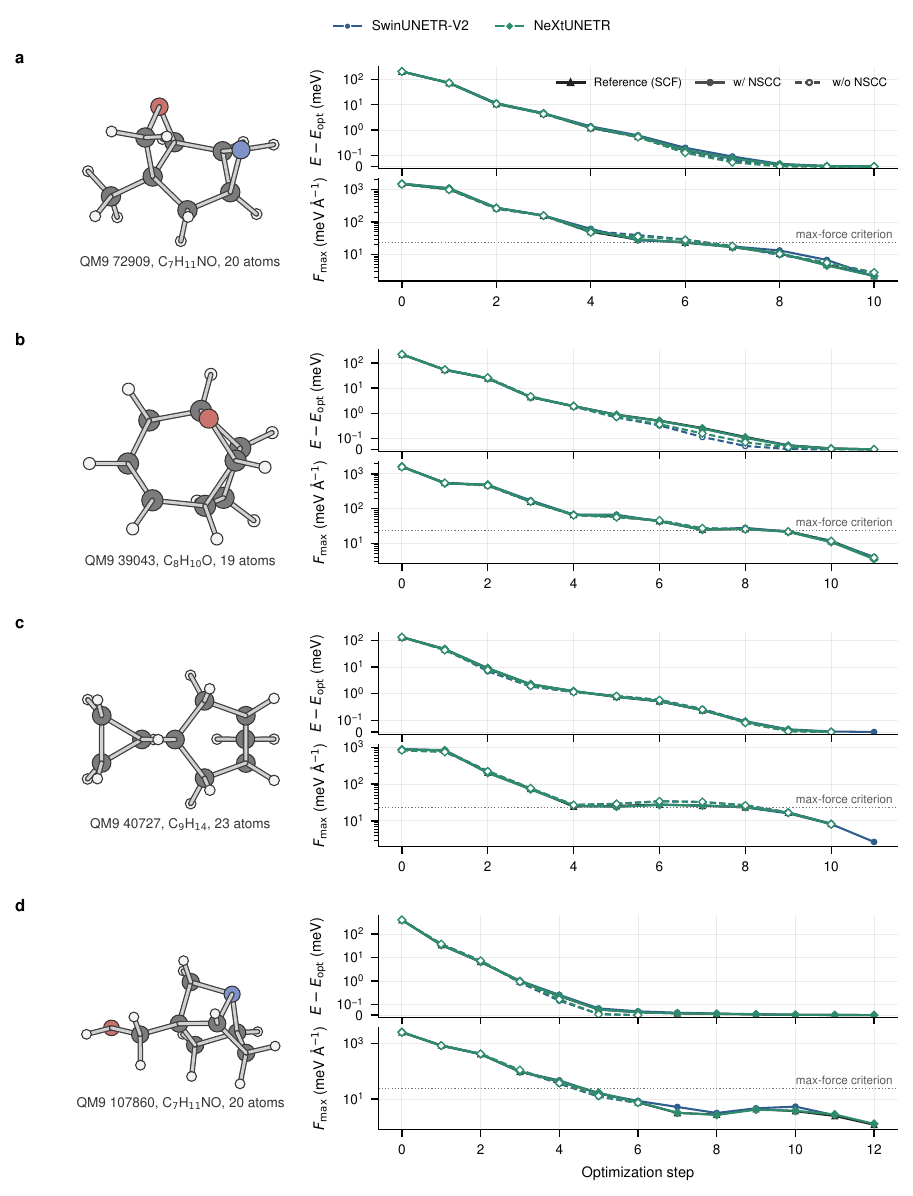}}
\end{center}
\small
\textbf{Supplementary Fig. 1 | Further geometry optimizations from MMFF94 starting structures.} Four molecules drawn at random, together with the example of Fig. 2d--f, from the molecules whose model optima all lie within 0.01~\AA{} RMSD of the optimum the reference reaches from the same start: QM9 72909 (C$_{7}$H$_{11}$NO; \textbf{a}), 39043 (C$_{8}$H$_{10}$O; \textbf{b}), 40727 (C$_{9}$H$_{14}$; \textbf{c}) and 107860 (C$_{7}$H$_{11}$NO; \textbf{d}). Left, reference optimum; right, energy relative to each surface's own optimum (top) and maximum force (bottom) along the optimization, for the reference SCF and for SwinUNETR-V2 and NeXtUNETR with and without the non-self-consistency correction (blue circles, SwinUNETR-V2; green diamonds, NeXtUNETR; black triangles, reference; solid, w/ NSCC; dashed, w/o NSCC). Energy on a symmetric logarithmic axis. Dotted: maximum-force criterion.

\normalsize

\clearpage
\begin{center}
\makebox[\linewidth][c]{\includegraphics[width=158.2mm,height=0.62\textheight,keepaspectratio]{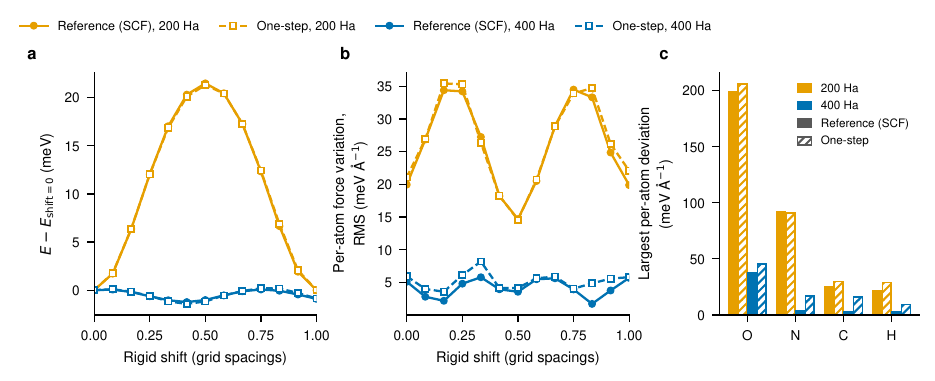}}
\end{center}
\small
\textbf{Supplementary Fig. 2 | The egg-box effect of the 200~Ha grid and the 400~Ha evaluation.} The relaxed alanine dipeptide (C7eq) translated rigidly by one spacing of the 200~Ha grid along each axis (0.144~\AA{} along the cell diagonal; 13 points). \textbf{a}, Energy relative to the unshifted geometry, for the reference SCF (solid lines, filled circles) and the one-step calculation (SwinUNETR-V2; dashed lines, open squares), evaluated on the 200~Ha grid (orange) and on the 400~Ha grid (blue); the 400~Ha curves are not periodic in the 200~Ha spacing. \textbf{b}, Per-atom force variation: deviation of each force component from its mean over the 13 points, net force removed at each point, RMS over atoms. \textbf{c}, Largest deviation of a per-atom force vector over the translation, by element (filled, SCF; hatched, one-step). One-step forces are w/ NSCC; in this CPU calculation the fixed-D term is evaluated on the 400~Ha grid but the Kohn--Sham matrix of the output density that enters the NSCC on the 200~Ha grid (the GPU implementation evaluates both on the 400~Ha grid; fully consistent, the 400~Ha RMS is 5.3 instead of 5.4~meV~\AA{}$^{-1}$).

\normalsize

\clearpage
\textbf{Supplementary Table 1 | Inference in fp32 and TF32.} Density NMAE (\%, mean over the 500 QM9 test molecules of Fig. 1) and network forward time per molecule, with TF32 tensor-core arithmetic disabled (fp32, used throughout the paper) and enabled; forward times on one NVIDIA H200 NVL GPU with batch size 1, as in Fig. 1d. ``Prediction change'' is the mean relative change of the predicted density between the two settings. All networks we trained used TF32 (PyTorch's default for cuDNN convolutions).

\begin{center}
\footnotesize
\setlength{\tabcolsep}{4pt}
\resizebox{\ifdim\width>\linewidth\linewidth\else\width\fi}{!}{%
\begin{tabular}{lrrrrrr}
\toprule
Model & Params (M) & NMAE fp32 & NMAE TF32 & Forward fp32 (ms) & Forward TF32 (ms) & Prediction change (\%) \\
\midrule
SwinUNETR-V2-18M & 18.5 & 0.369 & 0.370 & 88 & 76 & 0.006 \\
SwinUNETR-V2-73M & 73.1 & 0.255 & 0.255 & 145 & 94 & 0.004 \\
SwinUNETR-V2-292M & 291.7 & 0.204 & 0.204 & 278 & 116 & 0.003 \\
SwinUNETR-V2-656M & 655.9 & 0.215 & 0.215 & 550 & 165 & 0.003 \\
NeXtUNETR-17M & 17.5 & 0.187 & 0.188 & 38 & 33 & 0.010 \\
NeXtUNETR-67M & 67.5 & 0.133 & 0.133 & 74 & 60 & 0.008 \\
NeXtUNETR-150M & 150.1 & 0.116 & 0.116 & 115 & 89 & 0.007 \\
NeXtUNETR-265M & 265.4 & 0.120 & 0.120 & 157 & 115 & 0.007 \\
SR-ResNet-9M & 9.4 & 0.320 & 0.238 & 213 & 57 & 0.222 \\
SR-ResNet-70M & 70.2 & 0.198 & 0.191 & 1,544 & 208 & 0.123 \\
\bottomrule
\end{tabular}}
\end{center}

\clearpage
\textbf{Supplementary Table 2 | Single-point wall times.} Median time per molecule (s; three significant figures) over the 200 QM9 molecules of Extended Data Table 2 (4--25 atoms, median 19), each at its reference-SCF minimum. One calculation per molecule and row.

\begin{center}
\small
\resizebox{\ifdim\width>\linewidth\linewidth\else\width\fi}{!}{%
\begin{tabular}{lrrrr}
\toprule
 & A40 & A100 & H100 & CPU (16 cores) \\
\midrule
\multicolumn{5}{l}{\textbf{Reference SCF}} \\
energy & 10.6 & 2.29 & 2.17 & 61.8 \\
energy + forces & 12.8 & 2.69 & 2.41 & 68.4 \\
\multicolumn{5}{l}{\textbf{NeXtUNETR}} \\
network forward & 0.140 & 0.113 & 0.0686 & 1.89 \\
network backward & 0.159 & 0.128 & 0.0837 & 2.22 \\
energy & 1.63 & 0.543 & 0.451 & 6.95 \\
energy + forces w/o NSCC & 2.89 & 0.687 & 0.573 & 12.7 \\
energy + forces w/ NSCC & 4.39 & 0.997 & 0.801 & 24.2 \\
\multicolumn{5}{l}{\textbf{SwinUNETR-V2}} \\
network forward & 0.788 & 0.708 & 0.268 & 8.34 \\
network backward & 0.818 & 0.557 & 0.297 & 11.8 \\
energy & 2.32 & 1.15 & 0.664 & 13.1 \\
energy + forces w/o NSCC & 3.61 & 1.32 & 0.791 & 19.6 \\
energy + forces w/ NSCC & 5.78 & 2.03 & 1.23 & 39.2 \\
\bottomrule
\end{tabular}}
\end{center}

Times are for a new molecule in a running program: program start, library and model loading, file input and cell construction are excluded; every row starts from the same emptied caches.

Reference SCF: PBE, converged to 10$^{-11}$~Ha from the atomic guess, density and forces on the 400~Ha grid; median 16 SCF cycles (range 9--20). One-step: as described in Methods (network on the 200~Ha grid, energy and forces on the 400~Ha grid).

Networks run in fp32 (TF32 off). Network forward and backward: the network alone, from the SAD input. The backward pass gives the gradient with respect to the network input only (weights frozen); it is the network part of the w/ NSCC force.

Hardware: one NVIDIA A40 (48~GB), A100 SXM4 (80~GB) or H100 NVL GPU (GPU4PySCF 1.8.0 with PySCF 2.14.0, CuPy 14.1.1, PyTorch 2.3.1); CPU: 16 cores of an AMD EPYC 7543 (PySCF 2.6.2, PyTorch 2.3.1).

\end{document}